\documentclass{ws-ijmpe}
\usepackage[super,compress]{cite}
\usepackage{lineno,hyperref}
\usepackage{amsmath}
\usepackage{supertabular}
\usepackage{multirow}
\usepackage{amssymb}
\usepackage{amsmath}
\usepackage{graphicx}
\usepackage[normalem]{ulem}
\usepackage{longtable}
\usepackage{bm}
\usepackage{array}
\usepackage{etoolbox}
\usepackage{setspace}
\usepackage{color}
\usepackage{ulem}
\begin{document}

\markboth{Authors' Names}{Instructions for typing manuscripts (paper's title)}

\catchline{}{}{}{}{}
%%%%%%%%%%%%%%%%%%%%%%%%%%%%%%%%%%%%%%%%%%%%%%%%%%%%%%%%%%%%%%%%%%%%

\title{Predictions of $\alpha$ decay half-lives for even--even superheavy nuclei with $104 \leqslant Z \leqslant 128$  based on two--potential approach within cluster--formation model}

\author{Hong-Ming Liu}

\address{School of Nuclear Science and Technology, University of South China, 421001 Hengyang, People's Republic of China}

\author{Jun-Yao Xu}
\address{School of Nuclear Science and Technology, University of South China, 421001 Hengyang, People's Republic of China}

\author{Jun-Gang Deng}
\address{School of Nuclear Science and Technology, University of South China, 421001 Hengyang, People's Republic of China}

\author{Biao He}
\address{College of Physics and Electronics, Central South University, 410083 Changsha , People's Republic of China}

\author{Xiao-Hua Li}
\address{{School of Nuclear Science and Technology, University of South China, 421001 Hengyang, People's Republic of China}\\
lixiaohuaphysics@126.com}

\maketitle

\begin{history}
\received{Day Month Year}
\revised{Day Month Year}
%\accepted{Day Month Year}
%\comby{(xxxxxxxxxx)}
\end{history}

\begin{abstract}
 In present work, we systematically study the $\alpha$ decay half-lives of 170 even--even nuclei with $60 \leqslant Z \leqslant 118$ within the two--potential approach while the $\alpha$ decay preformation factor $P_\alpha$ is obtained by the cluster--formation model. The calculated results can well reproduce the experimental data. In addition, we extend this model to predict the $\alpha$ decay half-lives of 64 even--even nuclei with $104 \leqslant Z \leqslant 128$ whose $\alpha$ decay is energetically allowed or observed but not yet quantified. For comparing, the two famous models i.e. SemFIS proposed by D. Poenaru ${et\ al.}$ [\href {https://doi.org/10.1209/0295-5075/77/62001}{Europhys. Lett. \textbf{77} (2007) 62001}] and UDL proposed by C. Qi ${et\ al.}$ [\href {https://doi.org/10.1103/PhysRevLett.103.072501}{Phys. Rev. Lett. \textbf{103} (2009) 072501}] are used. The predicted results of  these models are basically consistent.  At the same time, through analyzing the changing trend of $\alpha$ decay energy $Q_{\alpha}$ of \emph{Z} = 118, 120, 122, 124, 126 and 128 isotopes nuclei with the increasing of neutron number \emph{N} and that of $\alpha$ decay preformation factor $P_\alpha$ of those isotopes even--even nuclei with the increasing of neutron number \emph{N}, \emph{N} = 178 may be a new neutron magic number.

\end{abstract}


\ccode{PACS numbers:}

%\tableofcontents

\section{INTRODUCTION}	

 In 1899, Rutherford discovered $\alpha$ decay mode. Later, Gurney and Condon \cite {1} and Gamow \cite {2} proposed the quantum tunneling theory to explain this process. Since then, $\alpha$ decay has become one of the most significant tools for acquiring nuclear structure information and identifying new isotopes or elements \cite {3,4,5,6,7,8,9,10}. In Gamow framework, the $\alpha$ decay process is described as a preformed $\alpha$ particle tunneling through $\alpha$--daughter nucleus potential barrier. The probability of $\alpha$ cluster formation in the parent nucleus i.e.  $\alpha$ decay preformation factor denotes as $P_\alpha$, which should be less than or equal to 1 \cite {11}. According to previous works, $P_\alpha$ can be  influenced by shell effect,  isospin asymmetry, unpaired nucleons of parent nucleus and so on \cite {11,12,13,14,15,16,17,18}. In theory, $P_\alpha$ can be obtained from the overlap between initial state and $\alpha$ decaying wave function describing the $\alpha$--core system \cite {19}. It also can be obtained from the initial tailored wave function of the parent nucleus in the $R$--matrix method \cite {20,21,22,23,24}. However, due to the complexity of the nuclear many--body problem and the uncertainty of the nuclear potential, it is very difficult to calculate the purely microcosmic ${P_\alpha}$. Experimentally, $\alpha$ preformation probabilities are usually extracted from rates of experimental $\alpha$ decay half-lives to theoretical  ones calculated without considering the preformation factors \cite{12,25}. In 2005, Xu and Ren systematically studied $\alpha$ decay half-lives of medium mass nuclei with $50 \leqslant Z \leqslant 82$ using the density--dependent cluster model (DDCM). Their results indicted that the $\alpha$ preformation factors are approximately alike for the same kinds of parent nuclei i.e. ${P_\alpha}$ = 0.43 for even--even nuclei, 0.35 for odd--\emph{A} nuclei, and 0.18 for doubly odd nuclei \cite{26}. Recently, we used the two--potential approach (TPA) \cite{27,28} to systematically study the behavior of ${P_\alpha}$ for odd--\emph{A} and odd--odd nuclei \cite{29,30}. Our results indicated that the ${P_\alpha}$ are smoothly changed with the mass number of parent nuclei for the same kind nuclei. In 2013, Ahmed ${et\ al.}$ proposed a new quantum--mechanical theory named cluster--formation model (CFM) \cite {31,32} to calculate the $\alpha$ preformation factors of even--even nuclei. For comparing, they also calculated the clustering amount (CA) of an $\alpha$ cluster and $P_\alpha$ for $^{212}$ Po. The results showed a good agreement with the $\alpha$ clustering value of Varga ${et\ al.}$ \cite{20,33} and the preformation factor value of Ni and Ren\cite{35} extracted by fitting the experimental data of $\alpha$ decay half-lives. Recently, Ahmed ${et\ al.}$ and Deng ${et\ al.}$ extended this model to the cases of odd--\emph{A} and odd--odd heavy nuclei \cite{14,37,38}. Their results can well reproduce experimental data.

Superheavy nuclei (SHN) present a constant challenge to experimental research due to their elusive synthesis and widely unknown properties. In the mid-1960s, SHN became a hot topic in nuclear physics\cite {16,37}. Great efforts have been done in the experiment and theory in this field, while the developments in heavy-ion beam technology have been also accelerating the experimental synthesis and theoretical research  \cite{40,41,42}.  For SHN, $\alpha$ decay as a dominant decay mode can provides abundant nuclear structure information such as the nuclear shell structure, properties of ground state and so on\cite {43,44,45,46,47,48,49,50,51,52}. Up to now, various theoretical models, such as cluster model \cite{53}, generalized liquid drop model \cite{54}, fission--like model \cite{55} and so on, have been used to study the $\alpha$ decay half-lives. These models can reproduce the experimental $\alpha$ decay half-lives well. The two--potential approach (TPA)\cite{27,28} was originally proposed to deal with quasi-stationary problem. More recently, it has been widely used to deal with $\alpha$ decay \cite {29,30,56,57,58,59,60,61}. In our previous works \cite {29,30,59,60,61}, we systematically studied the the $\alpha$ decay half-lives of even--even, odd--\emph{A} and doubly--odd nuclei by TPA. These results can well reproduce experimental data.

Recently, superheavy element \emph{Z} = 118 has been synthesized in the laboratory by $^{48}$Ca-induced hot fusion reactions with actinide Cf target \cite{3,7}. Many laboratories are trying to synthesize the new elements such as \emph{Z} = 119, \emph{Z} = 120 and so on. For providing a theoretical reference for the experiment, the aim of this work is to predict the $\alpha$ decay half-lives $T_{1/2}$ of even--even superheavy nuclei with $104 \leqslant Z \leqslant 128$ using TPA with CFM. For comparing, the predictions used two famous models i.e. the SemFIS \cite{62} based on fission theory and the UDL\cite{63} are also included.

The article is organized as follows. In next section, the theoretical frameworks of the TPA and CFM are briefly presented. The detailed results and discussion are presented in Section 3. Finally, a summary is given in Section 4.

\section{THEORETICAL FRAMEWORK}

\subsection{Two--potential approach}
     In the framework of the TPA \cite{27,28}, the $\alpha$ decay half-life $T_{1/2} $ can be calculated as

\begin{equation}
T_{1/2}=\frac{\hbar ln2}{\Gamma},
\end{equation}
where $\hbar$ and $\Gamma$ are the reduced Planck constant and decay width which is widely used in the spherical nuclei calculations of $\alpha$ decay, respectively. Recently, Soylu and Evlice studied the deformation effects on cluster decays of radium isotopes, their calculated results indicated that the deformation of the cluster is more important than the daughter\cite{64}, and the deformation of the daughter nucleus has little effect for the $\alpha$ decay half-lives of nuclei. Meanwhile, for the unified description, we don't consider the deformation effects. In this framework, the $\alpha$ decay width can be calculated as
\begin{equation}
\Gamma=\frac{\hbar^2 P_\alpha F P}{4 \mu},
\end{equation}
where $\mu$ = $\frac{m_d m_\alpha}{m_d+m_\alpha}$ is the reduced mass between $\alpha$ particle and the daughter nucleus with $m_d$ and $m_\alpha$ being the mass of the daughter nucleus and $\alpha$ particle, respectively. $P_{\alpha}$, the $\alpha$ decay preformation factor, is calculated by CFM whose more detailed information is given in the Sect.2.2.
\\ \emph{P} is the penetration probability, namely the Gamow factor, obtained by the WKB approximation \cite{61}. It can be expressed as
\begin{equation}
P=\rm{exp}(-2  \int_{r_2}^{r_3} k(r)\,dr),
\end{equation}
where k(r) = $\sqrt{\frac{2 \mu}{\hbar^2}{\left|Q_\alpha-V{(r)}\right|}}$ is the wave number of the $\alpha$ particle, \emph{r} is the center-of-mass distance between the preformed $\alpha$ particle and the daughter nucleus. $V(r)$ and $\emph{Q}_\alpha$ are $\alpha$--daughter nucleus interaction potential and $\alpha$ decay energy, respectively. \emph{F} is the normalized factor, representing the assault frequency, can be obtained by
\begin{equation}
F \int_{r_1}^{r_2} \frac{1}{2 k(r)}\,dr=1,
\end{equation}
 where $\emph{r}_1$, $\emph{r}_2$ and the above $\emph{r}_3$ are the classical turning points which satisfy the conditions $\emph{V}(r_1)$ = $V(r_2)$ = $V(r_3)$ = $\emph{Q}_\alpha$. The interaction potential between the preformed $\alpha$ particle and the daughter nucleus $V(r)$ is composed of the nuclear potential $V_{N}(r)$, Coulomb potential $V_{C}(r)$ and centrifugal potential $V_{l}(r)$. It can be expressed as
\begin{equation}
V(r) = V_{N}(r)+V_{C}(r)+V_{l}(r).
\end{equation}
  In this work, we choose a type of cosh parameterized form for the nuclear potential \cite{61}. It can be expressed as
\begin{equation}
V_{N}(r)=-V_0 \frac{1+ \rm{cosh}(\emph{R/a})}{\rm{cosh}(\emph{r/a})+ \rm{cosh}(\emph{R/a})},
\end{equation}
where $V_0$ and $a$ are the depth and diffuseness of nuclear potential, respectively. In our previous work \cite{59}, through analyzing the experimental $\alpha$ decay half-lives of 164 even--even nuclei, a set of isospin dependent parameters was obtained i.e., $a$ = 0.5958 fm and $V_0 = (192.42+31.059\frac{N_d-Z_d}{A_d}) \quad{\rm{MeV}}$ with $N_d$ , $Z_d$ and $A_d$ being the neutron, proton and mass number of the daughter nucleus, respectively. \emph{R}, the nuclear potential sharp radius, is parameterized as\cite{65}
\begin{equation}
R=1.28 A^{1/3}-0.76+0.8 A^{-1/3},
\end{equation}
where $A$ is the mass number of the parent nucleus.

The Coulomb potential $V_{C}(r)$, obtained under the assumption of a uniformly charged sphere with radius \emph{R}, is expressed as
\begin{equation}
\label{eq7}
\
V_C(r)=\left\{\begin{array}{llll}

\frac{Z_{d}Z_{\alpha}e^2}{2R}[3-(\frac{r}{R})^2], &r<R, \\

\frac{Z_{d}Z_{\alpha}e^2}{r}, &r>R,

\end{array}\right.
\end {equation}
where $Z_\alpha=2$ is proton numbers of the $\alpha$ particle.

$V_{l}(r)$ is the centrifugal potential. It can be expressed as
\begin{equation}
V_{l}(r)=\frac{ {\hbar}^2(l+\frac{1}{2})^2}{2 \mu r^2},
\end{equation}
for $l(l+1)\rightarrow(l+\frac{1}{2})^2$ being a necessary correction for one-dimensional problems\cite{66}.
 Here $l$ is the orbital angular momentum taken away by the emitted $\alpha$ particle. $l$ = 0 for favored $\alpha$ decays, while $l\ne0$ for unfavored decays. In the case of even--even nuclei $\alpha$ decay, $\emph{l}$ = 0.

\subsection{Cluster--formation model}

The new quantum--mechanical theory named cluster--formation model (CFM) was first put forward to calculate the $\alpha$ preformation factors $P_\alpha$ of even--even nuclei by Ahmed ${et\ al.}$ \cite{31,32}. Later, Deng ${et\ al.}$ and  Ahmed ${et\ al.}$ extend this model to odd--\emph{A} and odd--odd nuclei \cite{14,37,38}.
In this framework, the total state $\Psi$ of the parent nucleus is a linear combination of its $n$ possible clusterization states $\Psi_i$, which can be expressed as
\begin{equation}
 \Psi=\sum_{i=1}^n \ a_i   \Psi_i,
\end{equation}
 where $a_i$ is the superposition coefficient of $\Psi_i$. It can be expressed as
\begin{equation}
a_i=\int  {\Psi_i^*}\Psi  \,d\tau.
\end{equation}
 According to the orthogonality condition,
\begin{equation}
\sum_{i=1}^n \left| a_i \right|^2=1.
\end{equation}
The total wave function is an eigenfunction of the total
Hamiltonian \emph{H}, which can be expressed as
\begin{equation}
H=\sum_{i=1}^n H_i,
\end{equation}
where $H_i$ denotes the Hamiltonian for the $i$th clusterization state
$\Psi_i$. Owe to all the clusterizations describing the same
nucleus, they are assumed as sharing the same total energy
\emph{E} of the total wave function. Therefore, the total energy \emph{E} can be represented as
\begin{equation}
E=\sum_{i=1}^n \left| a_i \right|^2 E=\sum_{i=1}^n E_{fi},
\end{equation}
where $E_{fi}$ is the formation energy of cluster in the \emph{i}th
clusterization state $\Psi_i$. Hence, the $\alpha$ preformation factor can be calculated by
\begin{equation}
P_\alpha=\left| a_\alpha \right|^2= \frac{ E_{f\alpha}}  {E},
\end{equation}
where $a_\alpha$ represents the superposition coefficient of the $\alpha$ clusterization state $\Psi_\alpha$, $\emph{E}_{f\alpha}$ represents the formation energy of the $\alpha$ cluster, and \emph{E} represents actually composed of the formation energy (intrinsic energy) of the $\alpha$ cluster and the interaction energy between the $\alpha$ cluster and daughter nuclei. Within the CFM, for even--even nuclei, the $\alpha$ cluster--formation energy $E_{f\alpha}$ and total energy \emph{E} of a considered system can be expressed as \cite{68}
\begin{eqnarray}
\ E_{f\alpha}=&3 B(A,Z)+B (A-4,Z-2)\nonumber\\
&-2 B(A-1,Z-1)-2 B(A-1,Z),\\
\ E=&B (A,Z)-B (A-4,Z-2)
\end{eqnarray}
where $B(A,Z)$ is the binding energy of nucleus with the mass number \emph{$A$} and proton number \emph{$Z$}. In this work, the data of nuclei binding energies are taken from the latest evaluated atomic mass table AME2016 \cite{70,71} or WS3+ \cite{76} for the unsynthesized nuclei (the detailed discuss is given in the Section 3).

\section{RESULTS AND DISCUSSIONS}

 \,In our previous work, we systematically studied the $\alpha$ decay half-lives of nuclei around the closed shell of \emph{Z} = 82, \emph{N} = 126 within proximity potential 1977 formalism\cite{67} while preformation factor $P_{\alpha}$ is calculated by the CFM. The calculations can well reproduce the experimental data  \cite{68}. In this work, we combine CFM and TPA to systematically predict the $\alpha$ decay half-lives of even--even superheavy nuclei with $104 \leqslant Z \leqslant 128$. To verify whether this model can well reproduce the experimental data or not, we compare the experimental $\alpha$ decay half-lives with calculated ones within this model for 170 even--even nuclei with $60 \leqslant Z \leqslant 118$. The experimental data are taken from NUBA2016 \cite{69}. Meanwhile, for benchmark, the two famous models i.e. SemFIS and UDL, TPA with $P_{\alpha}$ = 0.43 taken from Ref.[26] are used. All the calculated results are given in Tab. 1. In this Table, the first four columns denote mass number of parent nucleus, proton number of parent nucleus, experimental $\alpha$ decay energy $Q_{\alpha}$ which is taken from the latest evaluated atomic mass table AME2016\cite{70,71} and $\alpha$ preformation factors calculated by CFM, respectively. The fifth column denotes experimental $\alpha$ decay half-life. 
 The last four columns denote the calculated $\alpha$ decay half-life using TPA with $P_{\alpha}$ evaluated by CFM, UDL, SemFIS and TPA with $P_{\alpha}$ = 0.43, which are denoted as $T_{1/2}^{\rm{calc1}}$, $T_{1/2}^{\rm{calc2}}$, $T_{1/2}^{\rm{calc3}}$ and $T_{1/2}^{\rm{calc4}}$, respectively. From this table we can find that the results calculated by our model can reproduce the experimental data as well as TPA with $P_{\alpha}$ = 0.43, SemFIS and UDL. It is shown that TPA with CFM is an effective tool for studying the  $\alpha$ decay half-life.

  For more on this, intuitively, we plot the difference between the theoretical calculations obtained by our model, TPA with $P_{\alpha}$ = 0.43, SemFIS and UDL and experimental data  with the logarithmic form in Fig. 1. As can be seen from this figure, we can find that the values of $log_{10}{T_{1/2}^{\rm{calc}}}-log_{10}{T_{1/2}^{\rm{exp}}}$ fluctuate around zero in the region of $140 \leqslant  A  \leqslant 220$ and $260 \leqslant A \leqslant  300$, while the values of $log_{10}{T_{1/2}^{\rm{calc}}}$ are much larger than the values of $log_{10}{T_{1/2}^{\rm{exp}}}$ in the region of $220\leqslant A \leqslant 260$. This may be caused by effect of shell closure \cite{37,72}. In this region, the neutron number of parent nucleus is located near the neutron magic number $N = 152$. The deviation between the experimental $\alpha$ decay half-lives and calculated ones is represented by $\sigma$ which is defined as $\sigma$ = $\sqrt{\sum{(log_{10}{T_{1/2}^{\rm{calc}}}-log_{10}{T_{1/2}^{\rm{exp}}})^2}/n}$. Using the data from Tab. 1, we can obtain $\sigma_1$ = 0.453, $\sigma_2$ = 0.384, $\sigma_3$ = 0.326, $\sigma_4$ = 0.394, which denote standard deviations between $T_{1/2}^{\rm{calc1}}$, $T_{1/2}^{\rm{calc2}}$, $T_{1/2}^{\rm{calc3}}$,  $T_{1/2}^{\rm{calc4}}$ and $T_{1/2}^{\rm{exp}}$, respectively. It means that these four models can consistently reproduce the experimental data.

\begingroup
\renewcommand*{\arraystretch}{1}
%\centering
\begin{longtable}{cccccccccc}
\caption{Calculations of  $\alpha$ decay half-lives and the $\alpha$ preformation factors of even--even nuclei with $60 \leqslant Z \leqslant 118$. The experimental  $\alpha$ decay half-lives are taken from the latest evaluated nuclear properties table NUBASE2016 \cite{69}, and the $\alpha$ decay energies are taken from the latest evaluated atomic mass table AME2016\cite{70,71}. The $\alpha$ preformation factors $P_{\alpha}$ are calculated within the CFM \cite{14,31,32,37,38}. }
\label{table 1} \\
\hline
\hline$A$&$Z$&$Q_{\alpha}(\rm{MeV})$&$P_{\alpha}$&$T_{1/2}^{\rm{exp}}\rm{(s)}$&$T_{1/2}^{\rm{calc1}}\rm{(s)}$&$T_{1/2}^{\rm{calc2}}\rm{(s)}$&$T_{1/2}^{\rm{calc3}}\rm{(s)}$&$T_{1/2}^{\rm{calc4}}\rm{(s)}$\\
\hline
\endfirsthead
\multicolumn{6}{c}%
{{\tablename\ \thetable{} -- (\emph{continued}) }} \\
\hline
\hline$A$&$Z$&$Q_{\alpha}(\rm{MeV})$&$P_{\alpha}$&$T_{1/2}^{\rm{exp}}\rm{(s)}$&$T_{1/2}^{\rm{calc1}}\rm{(s)}$&$T_{1/2}^{\rm{calc2}}\rm{(s)}$&$T_{1/2}^{\rm{calc3}}\rm{(s)}$&$T_{1/2}^{\rm{calc4}}\rm{(s)}$\\
\hline
\endhead
\hline
\hline \multicolumn{6}{r}{{(\emph{continued on next page})}} \\
\endfoot
\hline
 \hline
\endlastfoot
$144$&60&1.903&0.196& $7.23\times10^{23}$ & $3.88\times10^{23}$&$5.90\times10^{22}$&$2.24\times10^{23}$&$1.77\times10^{23}$	\\
$146$&62&	2.529&0.198&$2.15\times10^{15}$&	$8.03\times10^{15}$&$2.58\times10^{15}$&$5.00\times10^{15}$&$3.70\times10^{15}$\\
$	148	$	&	62	&	1.987 	&	0.195 	&	$	1.99\times10^{23}$	&	$	9.42\times10^{23}$	&	$	1.36\times10^{23}$	&	$	 5.83\times10^{23}$&	$	 4.28\times10^{23}$\\
$	148	$	&	64	&	3.272 	&	0.199 	&	$	2.24\times10^{9}$	&	$	5.12\times10^{9}$&	$	2.76\times10^{9}$	&	$	3.33\times10^{9}$&	 $	 2.36\times10^{9}$	 \\
$	150	$	&	64	&	2.807 	&	0.202 	&	$5.65\times10^{13}$	&	$	1.58\times10^{14}$ &	$	5.96\times10^{13}$	&	$	1.06\times10^{14}$	 &	$	 7.41\times10^{13}$ \\
$	152	$	&	64	&	2.205 	&	0.221 	&	$3.41\times10^{21}$	&	$	1.17\times10^{22}$	&	$	2.30\times10^{21}$	&	$	8.92\times10^{21}$	 &	$	 6.04\times10^{21}$ \\
$	150	$	&	66	&	4.351 	&	0.236 	&	$1.18\times10^{3}$	&	$	1.76\times10^{3}$	&	$	1.72\times10^{3}$	&	$	1.46\times10^{3}$&	 $	 9.63\times10^{2}$	 \\
$	152	$	&	66	&	3.727 	&	0.239 	&	$8.51\times10^{6}$	&	$	1.90\times10^{7}$	&	$	1.45\times10^{7}$	&	$	1.56\times10^{7}$	 &	$	 1.06\times10^{7}$ \\
$	154	$	&	66	&	2.945 	&	0.238 	&	$9.47\times10^{13}$ &	$	1.33\times10^{14}$	&	$	5.83\times10^{13}$	&	$	1.08\times10^{14}$	 &	$	 7.38\times10^{13}$ \\
$	152	$	&	68	&	4.935 	&	0.239 	&	$1.14\times10^{1}$ &	$	1.67\times10^{1}$	&	$	1.76\times10^{1}$	&	$	1.36\times10^{1}$	 &	$	 9.26\times10^{0}$ \\
$	154	$	&	68	&	4.280 	&	0.257 	&	$4.75\times10^{4}$ &	$	5.65\times10^{4}$	&	$	5.28\times10^{4}$	&	$	4.79\times10^{4}$	 &	$	 3.37\times10^{4}$ \\
$	156	$	&	68	&	3.481 	&	0.252 	&	$1.73\times10^{10}$ &	$	2.96\times10^{10}$	&	$	1.80\times10^{10}$	&	$	 2.40\times10^{10}$
&	$	 1.73\times10^{10}$	 \\
$	154	$	&	70	&	5.474 	&	0.231 	&	$4.42\times10^{-1}$ &	$	5.87\times10^{-1}$	&	$	6.06\times10^{-1}$	&	$	4.41\times10^{-1}$	 
&	$	3.15\times10^{-1}$ \\
$	156	$	&	70	&	4.809 	&	0.257 	&	$2.56\times10^{2}$  &	$	7.28\times10^{2}$	&	$	7.28\times10^{2}$	&	$	5.85\times10^{2}$&	 $	 4.35\times10^{2}$	 \\
$	158	$	&	70	&	4.170 	&	0.264 	&	$4.23\times10^{6}$ &	$	3.86\times10^{6}$	&	$	3.16\times10^{6}$	&	$	3.14\times10^{6}$	 &	$	 2.37\times10^{6}$ \\
$	156	$	&	72	&	6.025 	&	0.283 	&	$2.36\times10^{-2}$ &	$	2.43\times10^{-2}$	&	$	3.05\times10^{-2}$	&	$	 2.12\times10^{-2}$&	$	 1.60\times10^{-2}$	 \\
$	158	$	&	72	&	5.405 	&	0.243 	&	$2.23\times10^{0}$ &	$	1.01\times10^{1}$	&	$	9.92\times10^{0}$	&	$	7.22\times10^{0}$&	 $	 5.68\times10^{0}$	 \\
$	160	$	&	72	&	4.902 	&	0.255 	&	$1.89\times10^{3}$ &	$	2.56\times10^{3}$	&	$	2.37\times10^{3}$	&	$	1.93\times10^{3}$	 &	$	 1.52\times10^{3}$ \\
$	162	$	&	72	&	4.416 	&	0.241 	&	$4.86\times10^{5}$ &	$	1.53\times10^{6}$	&	$	1.14\times10^{6}$	&	$	1.11\times10^{6}$&	 $	 8.60\times10^{5}$	 \\
$	174	$	&	72	&	2.495 	&	0.144 	&	$6.31\times10^{22}$ &	$	3.16\times10^{24}$	&	$	3.31\times10^{23}$	&	$	 4.02\times10^{24}$&	$	 1.06\times10^{23}$	 \\
$	158	$	&	74	&	6.615 	&	0.251 	&	$1.25\times10^{-3}$ &	$	1.55\times10^{-3}$	&	$	1.68\times10^{-3}$	&	$	1.14\times10^{-3}$	 &	$	 9.07\times10^{-4}$ \\
$	160	$	&	74	&	6.065 	&	0.285 	&	$1.03\times10^{-1}$ &	$	1.28\times10^{-1}$	&	$	1.50\times10^{-1}$	&	$	1.02\times10^{-1}$	 &	$	 8.45\times10^{-2}$\\
$	162	$	&	74	&	5.678 	&	0.240 	&	$2.63\times10^{0}$ &	$	5.39\times10^{0}$	&	$	5.12\times10^{0}$	&	$	3.69\times10^{0}$	 &	$	 3.01\times10^{0}$ \\
$	164	$	&	74	&	5.278 	&	0.251 	&	$1.65\times10^{2}$ &	$	3.21\times10^{2}$	&	$	2.98\times10^{2}$	&	$	2.36\times10^{2}$	 &	$	 1.87\times10^{2}$ \\
$	166	$	&	74	&	4.856 	&	0.231 	&	$5.47\times10^{4}$ &	$	4.88\times10^{4}$	&	$	3.76\times10^{4}$	&	$	3.47\times10^{4}$&	 $	 2.62\times10^{4}$	 \\
$	168	$	&	74	&	4.501 	&	0.235 	&	$1.59\times10^{6}$ &	$	5.26\times10^{6}$	&	$	3.69\times10^{6}$	&	$	4.13\times10^{6}$	 
&	$	 3.88\times10^{6}$ \\
$	180	$	&	74	&	2.516 	&	0.162 	&	$5.68\times10^{25}$ &	$	5.54\times10^{25}$	&	$	5.82\times10^{24}$	&	$	1.23\times10^{26}$	 &	$	 2.09\times10^{25}$ \\
$	162	$	&	76	&	6.765 	&	0.247 	&	$	2.10 \times10^{-3}$	&	$	3.16\times10^{-3}$	&	$	3.18\times10^{-3}$	&	$	 2.10\times10^{-3}$&	$	 1.81\times10^{-3}$	\\
$	166	$	&	76	&	6.142 	&	0.243 	&	$	2.95 \times10^{-1}$	&	$	4.99\times10^{-1}$	&	$	4.80\times10^{-1}$	&	$	 3.43\times10^{-1}$	&	$	 2.82\times10^{-1}$\\
$	168	$	&	76	&	5.816 	&	0.247 	&	$	4.84 \times10^{0}$	&	$	9.75\times10^{0}$	&	$	9.26\times10^{0}$	&	$	    7.17\times10^{0}$&	$	 5.61\times10^{0}$	\\
$	170	$	&	76	&	5.537 	&	0.237 	&	$	7.75 \times10^{1}$	&	$	1.62\times10^{2}$	&	$	1.42\times10^{2}$	&	$	    1.23\times10^{2}$&	$	 8.94\times10^{1}$	\\
$	172	$	&	76	&	5.224 	&	0.239 	&	$	1.71 \times10^{3}$	&	$	4.84\times10^{3}$	&	$	4.03\times10^{3}$	&	$	    4.01\times10^{3}$&	$	 2.69\times10^{3}$	\\
$	174	$	&	76	&	4.871 	&	0.226 	&	$	1.78 \times10^{5}$	&	$	3.64\times10^{5}$	&	$	2.62\times10^{5}$	&	$	    3.18\times10^{5}$&	$	 1.91\times10^{5}$	\\
$	186	$	&	76	&	2.822 	&	0.157 	&	$	6.31 \times10^{22}$	&	$	9.81\times10^{22}$	&$1.33\times10^{22}$	&	$	  2.97\times10^{23}$&	$	 3.58\times10^{22}$	\\
$	166	$	&	78	&	7.285 	&	0.286 	&	$	3.00 \times10^{-4}$	&	$	3.36\times10^{-4}$	&$3.80\times10^{-4}$	&	$	  2.56\times10^{-4}$&	$	 2.23\times10^{-4}$	\\
$	172	$	&	78	&	6.463 	&	0.254 	&	$	1.00 \times10^{-1}$	&	$	1.76\times10^{-1}$	&$1.77\times10^{-1}$	&	$	  1.37\times10^{-1}$	&	$	 1.04\times10^{-1}$\\
$	174	$	&	78	&	6.183 	&	0.247 	&	$	1.16 \times10^{0}$	&	$	2.02\times10^{0}$	&	$ 1.92\times10^{0}$	&	$	    1.62\times10^{0}$&	$	 1.16\times10^{0}$	\\
$	176	$	&	78	&	5.885 	&	0.260 	&	$	1.57 \times10^{1}$	&	$	3.03\times10^{1}$	&	$ 2.96\times10^{1}$	&	$	    2.82\times10^{1}$	&	$	 1.83\times10^{1}$\\
$	178	$	&	78	&	5.572 	&	0.227 	&	$	2.68 \times10^{2}$	&	$	8.26\times10^{2}$	&	$ 6.72\times10^{2}$	&	$	    7.46\times10^{2}$&	$	 4.35\times10^{2}$	\\
$	180	$	&	78	&	5.237 	&	0.204 	&	$	1.87 \times10^{4}$	&	$	3.83\times10^{4}$	&	$ 2.62\times10^{4}$	&	$	    3.55\times10^{4}$	&	$	 1.82\times10^{4}$\\
$	182	$	&	78	&	4.951 	&	0.187 	&	$	4.19 \times10^{5}$	&	$	1.36\times10^{6}$	&	$ 7.96\times10^{5}$	&	$	    1.36\times10^{6}$	&	$	 5.92\times10^{5}$\\
$	184	$	&	78	&	4.599 	&	0.186 	&	$	5.86 \times10^{7}$	&	$	1.63\times10^{8}$	&	$ 8.44\times10^{7}$	&	$	    1.99\times10^{8}$&	$	 7.05\times10^{7}$	\\
$	186	$	&	78	&	4.320 	&	0.173 	&	$	5.35 \times10^{9}$	&	$	1.16\times10^{10}$	&$5.02\times10^{9}$	&	$	  1.67\times10^{10}$	 &	$	 4.69\times10^{9}$ \\
$	188	$	&	78	&	4.007 	&	0.216 	&	$	6.76 \times10^{12}$	&	$	1.81\times10^{12}$	&$8.35\times10^{11}$	&	$	  4.33\times10^{12}$&	$	 9.09\times10^{11}$	\\
$	190	$	&	78	&	3.269 	&	0.203 	&	$	2.05 \times10^{19}$	&	$	1.09\times10^{19}$	&$2.70\times10^{18}$	&	$	  4.11\times10^{19}$	&	$	 5.13\times10^{18}$\\
$	172	$	&	80	&	7.525 	&	0.290 	&	$	2.31 \times10^{-4}$	&	$	3.09\times10^{-4}$	&$3.42\times10^{-4}$	&	$	  2.44\times10^{-4}$
&	$	 2.08\times10^{-4}$	\\
$	174	$	&	80	&	7.233 	&	0.244 	&	$	2.00 \times10^{-3}$	&	$	2.72\times10^{-3}$	&$2.55\times10^{-3}$	&	$	  1.89\times10^{-3}$	
&	$	 1.54\times10^{-3}$\\
$	176	$	&	80	&	6.897 	&	0.267 	&	$	2.23 \times10^{-2}$	&	$	2.99\times10^{-2}$	&$3.07\times10^{-2}$	&	$	  2.41\times10^{-2}$	&	$	1.86\times10^{-2}$\\
$	178	$	&	80	&	6.577 	&	0.259 	&	$	2.98 \times10^{-1}$	&	$	4.01\times10^{-1}$	&$3.94\times10^{-1}$	&	$	  3.36\times10^{-1}$	&	$	 2.41\times10^{-1}$\\
$	180	$	&	80	&	6.259 	&	0.265 	&	$	5.37 \times10^{0}$	&	$	6.14\times10^{0}$	&	$ 6.05\times10^{0}$	&	$	    5.80\times10^{0}$&	$	 3.79\times10^{0}$	\\
$	182	$	&	80	&	5.996 	&	0.254 	&	$	7.80 \times10^{1}$	&	$	7.37\times10^{1}$	&	$	6.77\times10^{1}$	&	$	    7.48\times10^{1}$&	$	 4.35\times10^{1}$	\\
$	184	$	&	80	&	5.662 	&	0.243 	&	$	2.77 \times10^{3}$	&	$	2.28\times10^{3}$	&	$	1.91\times10^{3}$	&	$	    2.54\times10^{3}$&	$	 1.29\times10^{3}$	\\
$	186	$	&	80	&	5.204 	&	0.247 	&	$	5.02 \times10^{5}$	&	$	4.26\times10^{5}$	&	$	3.26\times10^{5}$	&	$	    5.71\times10^{5}$&	$	 2.44\times10^{5}$	\\
$	188	$	&	80	&	4.707 	&	0.239 	&	$	3.33 \times10^{9}$	&	$	3.21\times10^{8}$	&	$	2.02\times10^{8}$	&	$	    5.23\times10^{8}$	&	$	 1.79\times10^{8}$\\
$	180	$	&	82	&	7.419 	&	0.248 	&	$	4.10 \times10^{-3}$	&	$	3.52\times10^{-3}$	&	$	3.26\times10^{-3}$	&	$	 2.64\times10^{-3}$	
&	$	 2.03\times10^{-3}$\\
$	184	$	&	82	&	6.773 	&	0.236 	&	$	6.12 \times10^{-1}$	&	$	4.93\times10^{-1}$	&	$	4.31\times10^{-1}$	&	$	 4.17\times10^{-1}$	
&	$	 2.71\times10^{-1}$\\
$	186	$	&	82	&	6.470 	&	0.230 	&	$	1.18 \times10^{1}$	&	$	6.62\times10^{0}$	&	$	5.49\times10^{0}$	&	$	    6.07\times10^{0}$	
&	$	 3.55\times10^{0}$\\
$	188	$	&	82	&	6.109 	&	0.222 	&	$	2.68 \times10^{2}$	&	$	1.92\times10^{2}$	&	$	1.49\times10^{2}$	&	$	    1.97\times10^{2}$
&	$	 9.94\times10^{1}$	\\
$	190	$	&	82	&	5.697 	&	0.215 	&	$	1.76 \times10^{4}$	&	$	1.38\times10^{4}$	&	$	9.68\times10^{3}$	&	$	    1.62\times10^{4}$
&	$	 6.90\times10^{3}$	\\
$	192	$	&	82	&	5.221 	&	0.210 	&	$	3.52 \times10^{6}$	&	$	3.69\times10^{6}$	&	$	2.25\times10^{6}$	&	$	    5.26\times10^{6}$	
&	$	1.80\times10^{6}$\\
$	194	$	&	82	&	4.738 	&	0.198 	&	$	1.71 \times10^{10}$	&	$	2.68\times10^{9}$	&	$	1.31\times10^{9}$	&	$	    4.78\times10^{9}$
&	$	 1.23\times10^{9}$	\\
$	210	$	&	82	&	3.793 	&	0.107 	&	$	9.26 \times10^{16}$	&	$	3.47\times10^{16}$	&	$	6.26\times10^{15}$	&	$	  3.26\times10^{15}$	&	$	8.61\times10^{15}$\\
$	190	$	&	84	&	7.693 	&	0.262 	&	$	2.46 \times10^{-3}$	&	$	1.86\times10^{-3}$	&	$	1.84\times10^{-3}$	&	$	 1.75\times10^{-3}$
&	$	1.13\times10^{-3}$	\\
$	194	$	&	84	&	6.987 	&	0.235 	&	$	3.92 \times10^{-1}$	&	$	3.87\times10^{-1}$	&	$	3.41\times10^{-1}$	&	$	 3.80\times10^{-1}$
&	$	 2.11\times10^{-1}$	\\
$	196	$	&	84	&	6.658 	&	0.222 	&	$	5.67 \times10^{0}$	&	$	6.29\times10^{0}$	&	$	5.16\times10^{0}$	&	$	    6.51\times10^{0}$
&	$	 3.25\times10^{0}$	\\
$	198	$	&	84	&	6.310 	&	0.206 	&	$	1.85 \times10^{2}$	&	$	1.58\times10^{2}$	&	$	1.16\times10^{2}$	&	$	    1.73\times10^{2}$	
&	$	 7.56\times10^{1}$\\
$	200	$	&	84	&	5.981 	&	0.187 	&	$	6.20 \times10^{3}$	&	$	4.46\times10^{3}$	&	$	2.84\times10^{3}$	&	$	    5.19\times10^{3}$	
&	$	 1.93\times10^{3}$\\
$	202	$	&	84	&	5.700 	&	0.178 	&	$	1.39 \times10^{5}$&	  $	9.27\times10^{4}$	&	$	5.39\times10^{4}$&	$	    1.24\times10^{5}$
&	$	 3.83\times10^{4}$	 \\
$	204	$	&	84	&	5.485 	&	0.158 	&	$	1.88 \times10^{6}$&	  $	1.18\times10^{6}$	&	$	5.86\times10^{5}$&	$	    1.73\times10^{6}$	 
&	$	4.32\times10^{5}$ \\
$	206	$	&	84	&	5.327 	&	0.145 	&	$	1.39 \times10^{7}$&	  $	8.18\times10^{6}$	&	$	3.64\times10^{6}$&	$	    1.39\times10^{7}$	 &	$	 2.76\times10^{6}$ \\
$	208	$	&	84	&	5.216 	&	0.135 	&	$	9.15 \times10^{7}$	  &$3.35\times10^{7}$ &	$	1.36\times10^{7}$	  &	$   6.74\times10^{7}$ &	$	 1.05\times10^{7}$ 	 \\
$	210	$	&	84	&	5.408 	&	0.105 	&	$	1.20 \times10^{7}$	  &$3.37\times10^{6}$ &	$	1.15\times10^{6}$	  &	$   6.27\times10^{6}$
&	$	8.21\times10^{5}$ 	 \\
$	212	$	&	84	&	8.954 	&	0.221 	&	$	2.95 \times10^{-7}$	  &$2.51\times10^{-7}$ &	$	2.34\times10^{-7}$  &	$ 1.70\times10^{-7}$  	 &	$	 1.29\times10^{-7}$\\
$	214	$	&	84	&	7.834 	&	0.213 	&	$	1.64 \times10^{-4}$	  &$2.50\times10^{-4}$ &	$	2.42\times10^{-4}$  &	$ 1.43\times10^{-4}$  	 &	$	 1.24\times10^{-4}$\\
$	216	$	&	84	&	6.907 	&	0.205 	&	$	1.45 \times10^{-1}$	  &$2.85\times10^{-1}$ &	$	2.61\times10^{-1}$  &	$ 1.33\times10^{-1}$  	 &	$	 1.35\times10^{-1}$\\
$	218	$	&	84	&	6.115 	&	0.196 	&	$	1.86 \times10^{2}$	  &$4.22\times10^{2}$ &	$	  3.43\times10^{2}$	&	$   1.51\times10^{2}$ &	$	 1.92\times10^{2}$ 	 \\
$	194	$	&	86	&	7.862 	&	0.262 	&	$	7.80 \times10^{-4}$	  &$3.08\times10^{-3}$ &	$	2.89\times10^{-3}$  &	$ 3.23\times10^{-3}$  	 &	$	 1.88\times10^{-3}$\\
$	196	$	&	86	&	7.617 	&	0.257 	&	$	4.70 \times10^{-3}$	  &$1.67\times10^{-2}$ &	$	1.55\times10^{-2}$  &	$ 1.89\times10^{-2}$  	 &	$	 9.97\times10^{-3}$\\
$	200	$	&	86	&	7.043 	&	0.228 	&	$	1.17 \times10^{0}$	  &$1.44\times10^{0}$ &	$	  1.18\times10^{0}$	&	$   1.81\times10^{0}$ &	$	 7.61\times10^{-1}$ 	 \\
$	202	$	&	86	&	6.773 	&	0.213 	&	$	1.23 \times10^{1}$	  &$1.44\times10^{1}$ &	$	  1.10\times10^{1}$	&	$   1.97\times10^{1}$  &	 $	 7.14\times10^{0}$	 \\
$	204	$	&	86	&	6.547 	&	0.194 	&	$	1.03 \times10^{2}$	  &$1.14\times10^{2}$ &	$	  7.78\times10^{1}$	&	$   1.67\times10^{2}$ &	$	 3.13\times10^{1}$ 	 \\
$	206	$	&	86	&	6.384 	&	0.181 	&	$	5.46 \times10^{2}$	  &$5.29\times10^{2}$ &	$	  3.35\times10^{2}$	&	$   8.68\times10^{2}$ &	$	 2.23\times10^{2}$ 	 \\
$	208	$	&	86	&	6.260 	&	0.163 	&	$	2.33 \times10^{3}$	  &$1.82\times10^{3}$ &	$	  1.04\times10^{3}$	&	$   3.27\times10^{3}$ &	$	 6.90\times10^{2}$ 	 \\
$	210	$	&	86	&	6.159 	&	0.152 	&	$	8.99 \times10^{3}$	  &$4.98\times10^{3}$ &	$	  2.63\times10^{3}$	&	$   1.02\times10^{4}$ &	$	 1.75\times10^{3}$ 	 \\
$	212	$	&	86	&	6.385 	&	0.121 	&	$	1.43 \times10^{3}$	  &$5.77\times10^{2}$ &	$	  2.58\times10^{2}$	&	$   1.11\times10^{3}$  &	 $	 1.62\times10^{2}$	 \\
$	214	$	&	86	&	9.208 	&	0.228 	&	$	2.70 \times10^{-7}$	  &$2.90\times10^{-7}$ &	$	2.61\times10^{-7}$  &	$ 1.74\times10^{-7}$  	 &	$	 1.54\times10^{-7}$\\
$	216	$	&	86	&	8.198 	&	0.237 	&	$	4.50 \times10^{-5}$	  &$1.12\times10^{-4}$ &	$	1.12\times10^{-4}$  &	$ 6.68\times10^{-5}$  	 &	$	6.15\times10^{-5}$\\
$	218	$	&	86	&	7.263 	&	0.234 	&	$	3.38 \times10^{-2}$	  &$9.21\times10^{-2}$ &	$	9.21\times10^{-2}$  &	$ 5.19\times10^{-2}$  	 &	$	 5.01\times10^{-2}$\\
$	220	$	&	86	&	6.405 	&	0.221 	&	$	5.56 \times10^{1}$	  &$1.74\times10^{2}$ &	$	  1.53\times10^{2}$	&	$   8.37\times10^{1}$ &	$	 8.94\times10^{1}$ 	 \\
$	222	$	&	86	&	5.591 	&	0.222 	&	$	3.30 \times10^{5}$	  &$1.08\times10^{6}$ &	$	  8.18\times10^{5}$	&	$   4.28\times10^{5}$  &	 $	 5.55\times10^{5}$ \\
$	202	$	&	88	&	7.880 	&	0.248 	&	$	4.10 \times10^{-3}$	  &$1.26\times10^{-2}$ &	$	1.10\times10^{-2}$  &	$ 1.79\times10^{-2}$  	 &	$	 7.27\times10^{-3}$\\
$	204	$	&	88	&	7.637 	&	0.237 	&	$	6.00 \times10^{-2}$	  &$7.18\times10^{-2}$ &	$	6.06\times10^{-2}$  &	$ 1.11\times10^{-1}$  	 &	$	 3.97\times10^{-2}$\\
$	208	$	&	88	&	7.273 	&	0.199 	&	$	1.27 \times10^{0}$	  &$1.24\times10^{0}$ &	$	  8.82\times10^{-1}$  &	$ 2.19\times10^{0}$  &	$	 5.71\times10^{-1}$	 \\
$	214	$	&	88	&	7.273 	&	0.139 	&	$	2.44 \times10^{0}$	  &$1.31\times10^{0}$ &	$	  6.83\times10^{-1}$  &	$ 2.67\times10^{0}$  	&	 $	 4.21\times10^{-1}$ \\
$	216	$	&	88	&	9.526 	&	0.239 	&	$	1.82 \times10^{-7}$	  &$2.35\times10^{-7}$ &	$	2.06\times10^{-7}$  &	$ 1.18\times10^{-7}$  	 &	$	 1.31\times10^{-7}$\\
$	218	$	&	88	&	8.546 	&	0.242 	&	$	2.52 \times10^{-5}$	  &$6.30\times10^{-5}$ &	$	6.04\times10^{-5}$  &	$ 3.28\times10^{-5}$  	 &	$	 3.54\times10^{-5}$\\
$	220	$	&	88	&	7.592 	&	0.240 	&	$	1.79 \times10^{-2}$	  &$4.45\times10^{-2}$ &	$	4.31\times10^{-2}$  &	$ 2.36\times10^{-2}$  	 &	$	 2.48\times10^{-2}$\\
$	222	$	&	88	&	6.678 	&	0.199 	&	$	3.36 \times10^{1}$	  &$1.13\times10^{2}$ &	$	  8.53\times10^{1}$	&	$   4.98\times10^{1}$  &	 $	 5.22\times10^{1}$	 \\
$	224	$	&	88	&	5.789 	&	0.184 	&	$	3.14 \times10^{5}$	  &$1.23\times10^{6}$ &	$	  7.34\times10^{5}$	&	$   4.73\times10^{5}$ &	$	 5.25\times10^{5}$ 	 \\
$	226	$	&	88	&	4.871 	&	0.182 	&	$	5.05 \times10^{10}$	  &$2.43\times10^{11}$ &	$	  1.06\times10^{11}$  &	$ 7.84\times10^{10}$  	 &	$	 1.02\times10^{11}$\\
$	212	$	&	90	&	7.958 	&	0.205 	&	$	3.17\times10^{-2}$	&	$	3.47\times10^{-2}$	&	$	2.50\times10^{-2}$	&	$	 7.30\times10^{-2}$	&	$	1.65\times10^{-2}$\\
$	214	$	&	90	&	7.827 	&	0.196 	&	$	8.70\times10^{-2}$	&	$	8.63\times10^{-2}$	&	$	6.02\times10^{-2}$	&	$	 2.06\times10^{-1}$	&	$	 3.94\times10^{-2}$\\
$	216	$	&	90	&	8.072 	&	0.159 	&	$	2.60\times10^{-2}$	&	$	1.60\times10^{-2}$&	  $	9.28\times10^{-2}$	    &	$	 3.57\times10^{-2}$	&	$	5.91\times10^{-3}$\\
$	218	$	&	90	&	9.849 	&	0.251 	&	$	1.17\times10^{-7}$	&	$	1.89\times10^{-7}$	&	$	1.60\times10^{-7}$	&	$	 7.24\times10^{-8}$&	$	 1.10\times10^{-7}$	\\
$	220	$	&	90	&	8.953 	&	0.247 	&	$	9.70\times10^{-6}$	&	$	2.60\times10^{-5}$	&	$	2.37\times10^{-5}$	&	$	 1.08\times10^{-5}$&	$	 1.49\times10^{-5}$	\\
$	222	$	&	90	&	8.127 	&	0.231 	&	$	2.24\times10^{-3}	$  &	$	5.54\times10^{-3}$	&	$	4.88\times10^{-3}	$  &	$	 2.42\times10^{-3}	$&	$	 2.98\times10^{-3}$\\
$	224	$	&	90	&	7.299 	&	0.199 	&	$	1.04\times10^{0}$   &	$	3.35\times10^{0}$	&	$	2.50\times10^{0}$	&	$	 1.40\times10^{0}$&	$	 1.55\times10^{0}$	\\
$	226	$	&	90	&	6.453 	&	0.182 	&	$	1.84\times10^{3}$	&	$	7.97\times10^{3}$	&	$	4.97\times10^{3}$	&	$	 3.25\times10^{3}$	&	$	 3.38\times10^{3}$	\\
$	228	$	&	90	&	5.520 	&	0.183 	&	$	6.03\times10^{7}$	&	$	3.08\times10^{8}$	&	$	1.57\times10^{ 8}$	&	$	 1.26\times10^{8}$&	$	 1.31\times10^{8}$	\\
$	230	$	&	90	&	4.770 	&	0.183 	&	$	2.38\times10^{12}$	&	$	1.44\times10^{13}$	&	$	5.43\times10^{12}$	&	$	 5.40\times10^{12}$&	$	 6.10\times10^{12}$	\\
$	232	$	&	90	&	4.082 	&	0.160 	&	$	4.42\times10^{17}$	&	$	4.15\times10^{18}$	&	$	9.18\times10^{17}$	&	$	 1.07\times10^{18}$	&	$	 1.54\times10^{18}$\\
$	216	$	&	92	&	8.530 	&	0.215 	&	$	6.90\times10^{-3}$	&	$	3.16\times10^{-3}$	&	$	2.29\times10^{-3}$	&	$	 8.04\times10^{-3}$	&	$	 1.58\times10^{-3}$\\
$	218	$	&	92	&	8.775 	&	0.189 	&	$	5.50\times10^{-4}$	&	$	6.54\times10^{-4}$	&	$	4.22\times10^{-4}$	&	$	 1.67\times10^{-3}$	&	$	 2.88\times10^{-4}$\\
$	222	$	&	92	&	9.478 	&	0.246 	&	$	4.70\times10^{-6}$	&	$	6.02\times10^{-6}$	&	$	5.02\times10^{-6}$	&	$	 1.85\times10^{-6}$	&	$	 3.44\times10^{-6}$\\
$	224	$	&	92	&	8.628 	&	0.228 	&	$	3.96\times10^{-4}$	&	$	1.05\times10^{-3}$	&	$	8.51\times10^{-4}$	&	$	 3.45\times10^{-4}$	&	$	 5.55\times10^{-4}$\\
$	226	$	&	92	&	7.701 	&	0.206 	&	$	2.69\times10^{-1}$	&	$	8.17\times10^{-1}$	&	$	6.04\times10^{-1}$	&	$	 2.83\times10^{-1}$	&	$	 3.92\times10^{-1}$\\
$	230	$	&	92	&	5.992 	&	0.187 	&	$	1.75\times10^{6}$	&	$	9.23\times10^{16}$	&	$	4.97\times10^{6}$	&	$	 3.87\times10^{6}$	&	$	 4.02\times10^{6}$\\
$	232	$	&	92	&	5.414 	&	0.169 	&	$	2.17\times10^{9}$	&	$	1.36\times10^{10}$	&	$	5.59\times10^{9}$	&	$	 5.69\times10^{9}$	&	$	 5.32\times10^{9}$\\
$	234	$	&	92	&	4.858 	&	0.150 	&	$	7.75\times10^{12}$	&	$	5.21\times10^{13}$	&	$	1.53\times10^{13}$	&	$	 1.94\times10^{13}$	&	$	 1.82\times10^{13}$\\
$	236	$	&	92	&	4.573 	&	0.153 	&	$	7.39\times10^{14}$	&	$	5.62\times10^{15}$	&	$	1.47\times10^{15}$	&	$	 1.85\times10^{15}$&	$	 2.00\times10^{15}$	\\
$	238	$	&	92	&	4.270 	&	0.137 	&	$	1.41\times10^{17}$	&	$	1.60\times10^{18}$	&	$	3.17\times10^{17}$	&	$	 3.41\times10^{17}$&	$	 5.12\times10^{17}$	\\
$	228	$	&	94	&	7.940 	&	0.230 	&	$	2.10\times10^{0}$	&	$	7.18\times10^{-1}$	&	$	5.56\times10^{-1}$	&	$	 1.88\times10^{-1}$&	$	 3.84\times10^{-1}$	\\
$	230	$	&	94	&	7.180 	&	0.200 	&	$	1.02\times10^{2}$	&	$	4.12\times10^{2}$	&	$	2.66\times10^{2}$	&	$	1.21\times10^{ 2}$&	$	 1.92\times10^{2}$	\\
$	232	$	&	94	&	6.716 	&	0.164 	&	$	1.74\times10^{4}$	&	$	3.68\times10^{4}$	&	$	1.85\times10^{4}$	&	$	1.19\times10^{ 4}$	&	$	 1.41\times10^{4}$\\
$	234	$	&	94	&	6.310 	&	0.152 	&	$	5.28\times10^{5}$	&	$	2.54\times10^{6}$	&	$	1.11\times10^{6}$	&	$	9.62\times10^{ 5}$	&	$	 8.97\times10^{5}$\\
$	236	$	&	94	&	5.867 	&	0.140 	&	$	9.02\times10^{7}$	&	$	4.33\times10^{8}$	&	$	1.58\times10^{8}$	&	$	1.75\times10^{ 8}$&	$	 1.41\times10^{8}$	\\
$	238	$	&	94	&	5.594 	&	0.145 	&	$	2.77\times10^{9}$	&	$	1.25\times10^{10}$	&	$	4.41\times10^{9}$	&	$	 5.50\times10^{9}$	&	$	 4.22\times10^{9}$\\
$	240	$	&	94	&	5.256 	&	0.129 	&	$	2.07\times10^{11}$	&	$	1.39\times10^{12}$	&	$	3.91\times10^{11}$	&	$	 4.96\times10^{11}$&	$	 4.19\times10^{11}$	\\
$	242	$	&	94	&	4.984 	&	0.127 	&	$	1.18\times10^{13}$	&	$	7.94\times10^{13}$	&	$	1.99\times10^{13}$	&	$	 2.17\times10^{13}$&	$	 2.35\times10^{13}$	\\
$	244	$	&	94	&	4.666 	&	0.133 	&	$	2.52\times10^{15}$	&	$	1.40\times10^{16}$	&	$	3.14\times10^{15}$	&	$	 2.54\times10^{15}$	&	$	 4.31\times10^{15}$\\
$	234	$	&	96	&	7.366 	&	0.191 	&	$	1.93\times10^{2}$	&	$	5.24\times10^{2}$	&	$	3.09\times10^{2}$	&	$	 1.51\times10^{2}$	&	$	 2.32\times10^{2}$\\
$	236	$	&	96	&	7.067 	&	0.160 	&	$	2.24\times10^{3}$	&	$	8.68\times10^{3}$	&	$	4.20\times10^{3}$	&	$	 2.95\times10^{3}$	&	$	 3.22\times10^{3}$\\
$	238	$	&	96	&	6.670 	&	0.137 	&	$	2.06\times10^{5}$	&	$	4.63\times10^{5}$	&	$	1.82\times10^{5}$	&	$	 1.74\times10^{5}$	&	$	 1.47\times10^{5}$\\
$	240	$	&	96	&	6.398 	&	0.137 	&	$	2.33\times10^{6}$	&	$	7.67\times10^{6}$	&	$	2.89\times10^{6}$	&	$	 3.44\times10^{6}$&	$	 2.45\times10^{6}$	\\
$	242	$	&	96	&	6.216 	&	0.122 	&	$	1.41\times10^{7}$	&	$	6.01\times10^{7}$	&	$	1.98\times10^{7}$	&	$	 2.61\times10^{7}$&	$	 1.71\times10^{7}$	\\
$	244	$	&	96	&	5.902 	&	0.144 	&	$	5.71\times10^{8}$	&	$	1.99\times10^{9}$	&	$	7.24\times10^{8}$	&	$	 9.70\times10^{8}$	&	$	 6.68\times10^{8}$\\
$	246	$	&	96	&	5.475 	&	0.142 	&	$	1.49\times10^{11}$	&	$	5.15\times10^{11}$	&	$	1.62\times10^{11}$	&	$	 2.02\times10^{11}$&	$	 1.70\times10^{11}$	\\
$	248	$	&	96	&	5.162 	&	0.146 	&	$	1.20\times10^{13}$	&	$	4.49\times10^{13}$	&	$	1.30\times10^{13}$	&	$	 1.25\times10^{13}$&	$	 1.53\times10^{13}$	\\
$	250	$	&	96	&	5.170 	&	0.144 	&	$	1.46\times10^{12}$	&	$	3.64\times10^{13}$	&	$	1.06\times10^{13}$	&	$	 5.97\times10^{12}$&	$	 1.22\times10^{13}$	\\
$	238	$	&	98	&	8.130 	&	0.183 	&	$	1.06\times10^{1}$	&	$	5.77\times10^{0}$	&	$	3.28\times10^{0}$	&	$	 1.82\times10^{0}$&	$	 2.46\times10^{0}$	\\
$	240	$	&	98	&	7.711 	&	0.156 	&	$	4.09\times10^{1}$	&	$	1.80\times10^{2}$	&	$	8.62\times10^{1}$	&	$	 6.64\times10^{1}$	&	$	 6.55\times10^{1}$\\
$	242	$	&	98	&	7.517 	&	0.150 	&	$	2.61\times10^{2}$	&	$	9.02\times10^{2}$	&	$	4.12\times10^{2}$	&	$	 4.19\times10^{2}$&	$	 3.14\times10^{2}$	\\
$	244	$	&	98	&	7.329 	&	0.153 	&	$	1.16\times10^{3}$	&	$	4.30\times10^{3}$	&	$	1.99\times10^{3}$	&	$	 2.46\times10^{3}$&	$1.53\times10^{3}$	\\
$	246	$	&	98	&	6.862 	&	0.155 	&	$	1.29\times10^{5}$	&	$	3.41\times10^{5}$	&	$	1.51\times10^{5}$	&	$	 2.14\times10^{5}$&	$	 1.23\times10^{5}$	\\
$	248	$	&	98	&	6.361 	&	0.138 	&	$	2.88\times10^{7}$	&	$	7.35\times10^{7}$	&	$	2.67\times10^{7}$	&	$	 4.04\times10^{7}$	&	$	 2.35\times10^{7}$\\
$	250	$	&	98	&	6.129 	&	0.136 	&	$	4.13\times10^{8}$	&	$	1.03\times10^{9}$	&	$	3.52\times10^{8}$	&	$	 4.79\times10^{8}$&	$	 3.27\times10^{8}$	\\
$	252	$	&	98	&	6.217 	&	0.146 	&	$	8.61\times10^{7}$	&	$	3.05\times10^{8}$	&	$	1.17\times10^{8}$	&	$	 1.15\times10^{8}$&	$	 1.04\times10^{8}$	\\
$	254	$	&	98	&	5.926 	&	0.154 	&	$	1.68\times10^{9}$	&	$	9.07\times10^{9}$	&	$	3.46\times10^{9}$	&	$	 2.32\times10^{9}$	&	$	 3.25\times10^{9}$\\
$	256	$	&	98	&	5.555 	&	1.037 	&	$	1.19\times10^{11}$	&	$	1.71\times10^{11}$	&	$	3.94\times10^{11}$	&	$	 1.52\times10^{11}$	&	$	 4.13\times10^{11}$\\
$	244	$	&	100	&	8.554 	&	0.171 	&	$	7.80\times10^{-1}$	&	$	1.23\times10^{0}$	&	$	6.40\times10^{-1}$	&	$	 5.57\times10^{-1}$	&	$	 4.90\times10^{-1}$\\
$	248	$	&	100	&	7.995 	&	0.164 	&	$	3.61\times10^{1}$	&	$	7.83\times10^{1}$	&	$	3.92\times10^{1}$	&	$	 5.33\times10^{1}$	&	$	 2.98\times10^{1}$\\
$	252	$	&	100	&	7.153 	&	0.154 	&	$	9.14\times10^{4}$	&	$	1.19\times10^{5}$	&	$	5.27\times10^{4}$	&	$	 8.45\times10^{4}$&	$	 4.26\times10^{4}$	\\
$	254	$	&	100	&	7.308 	&	0.135 	&	$	1.17\times10^{4}$	&	$	2.84\times10^{4}$	&	$	1.15\times10^{4}$	&	$	 1.59\times10^{4}$	&	$	 8.93\times10^{3}$\\
$	256	$	&	100	&	7.028 	&	0.154 	&	$	1.16\times10^{5}$	&	$	3.35\times10^{5}$	&	$	1.51\times10^{5}$	&	$	 1.71\times10^{5}$&	$	 1.20\times10^{5}$	\\
$	254	$	&	102	&	8.226 	&	0.140 	&	$	5.68\times10^{1}$	&	$	7.23\times10^{1}$	&	$	3.03\times10^{1}$	&	$	 5.15\times10^{1}$&	$	 2.36\times10^{1}$	\\
$	256	$	&	102	&	8.581 	&	0.153 	&	$	2.91\times10^{0}$	&	$	3.96\times10^{0}$	&	$	1.87\times10^{0}$	&	$	 3.16\times10^{0}$&	$	 1.41\times10^{0}$	\\
$	258	$	&	102	&	8.153 	&	0.156 	&	$	1.20\times10^{2}$	&	$	9.71\times10^{1}$	&	$	4.64\times10^{1}$	&	$	 7.31\times10^{1}$	&	$	 3.52\times10^{1}$\\
$	256	$	&	104	&	8.926 	&	0.156 	&	$	2.07\times10^{0}$	&	$	1.94\times10^{0}$	&	$	8.67\times10^{-1}$	&	$	 1.46\times10^{0}$	&	$	 7.02\times10^{-1}$\\
$	258	$	&	104	&	9.192 	&	0.174 	&	$	1.05\times10^{-1}$	&$2.51\times10^{-1}$	&	$	1.26\times10^{-1}$	&	$	2.40\times10^{-1}$	 &	$	 1.02\times10^{-1}$ \\
$	260	$	&	104	&	8.903 	&	0.162 	&	$	1.05\times10^{0}$	&	$	1.81\times10^{0}$	&	$	8.70\times10^{1}$	&	$	 1.70\times10^{0}$	&	$	 6.84\times10^{-1}$\\
$	260	$	&	106	&	9.901 	&	0.171 	&	$	1.23\times10^{-2}$	&$1.28\times10^{-2}$	&	$	5.97\times10^{-3}$	&	$	 1.15\times10^{-2}$&	$	 5.12\times10^{-3}$	 \\
$	264	$	&	108	&	10.590 &	0.176 	&	$	1.08\times10^{-3}$	&$8.95\times10^{-4}$	&	$	3.98\times10^{-4}$	&	$	 8.74\times10^{-4}$&	$	 3.67\times10^{-4}$	 \\
$	268	$	&	108	&	9.625 	&	0.191 	&	$	1.42\times10^{0}$	&	$	2.45\times10^{-1}$	&	$	1.27\times10^{-1}$	&	$	 3.26\times10^{-1}$&	$	1.09\times10^{-1}$	\\
$	270	$	&	108	&	9.065 	&	0.163 	&	$	9.00\times10^{0}$	&	$	1.25\times10^{1}$	&	$	5.57\times10^{0}$	&	$	 1.42\times10^{1}$	&	$	 4.73\times10^{0}$\\
$	270	$	&	110	&	11.115 	&	0.187 	&	$	2.05\times10^{-4}$	&$1.70\times10^{-4}$	&	$	7.52\times10^{-5}$&	$2.15 \times10^{-4}$	&	 $	 7.40\times10^{-5}$ \\
$	286	$	&	114	&	10.365 	&	0.177 	&	$	3.50\times10^{-1}$	&$1.49\times10^{-1}$	&	$	6.45\times10^{-2}$&	$2.02 \times10^{-1}$	&	 $	 6.14\times10^{-2}$ \\
$	288	$	&	114	&	10.065 	&	0.163 	&	$	7.50\times10^{-1}$	&$9.94\times10^{-1}$	&	$	4.04\times10^{-1}$&	$9.94 \times10^{-1}$	&	 $	 3.77\times10^{-1}$ \\
$	290	$	&	116	&	11.005 	&	0.175 	&	$	  8.00\times10^{-3}$	&$1.36\times10^{-2}$	&	$	5.40\times10^{-3}$&	$2.0 \times10^{-2}$	&	 $	 5.52\times10^{-3}$ \\
$	292	$	&	116	&	10.775 	&	0.171 	&	$	2.40\times10^{-2}$	&$4.84\times10^{-2}$	&	$	1.94\times10^{-2}$&	$5.82 \times10^{-2}$	&	 $	 1.93\times10^{-2}$ \\
$	294	$	&	118	&	11.835 	&	0.179 	&	$	  1.15\times10^{-3}$	&$5.39\times10^{-4}$	&	$	1.99\times10^{-4}$&	$8.89\times10^{-4}$	&	 $	 2.24\times10^{-4}$ \\

\end{longtable}
\endgroup

\begin{figure}
\centering
\includegraphics[width=15.0cm]{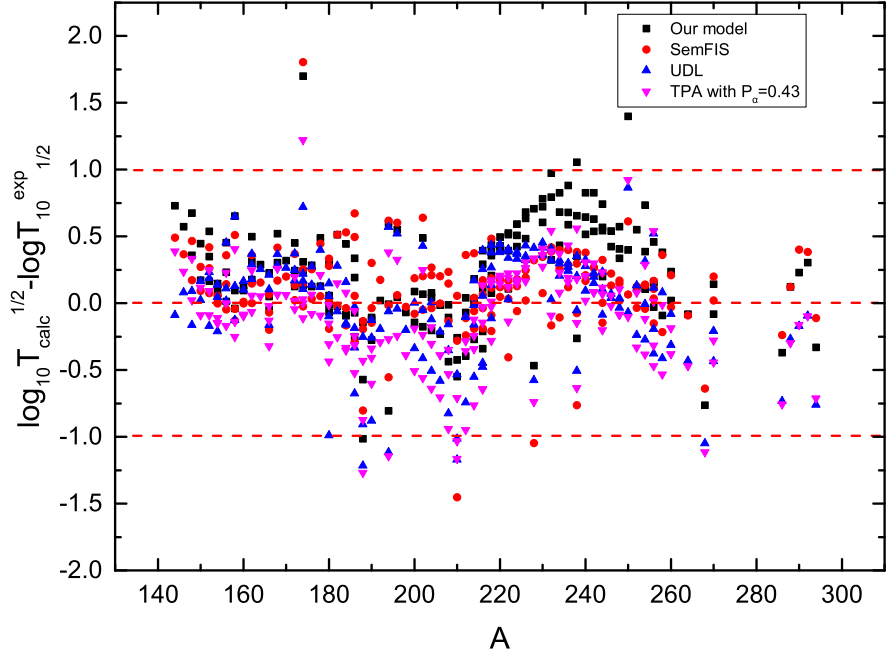}
\caption{(color online) The deviations of logarithmic form of $\alpha$ decay half-lives between calculations using four different models and experimental data as a function of neutron numbers. The black square, red circle, blue up triangle and magenta down triangle represent the deviations of our model, SemFIS, UDL and TPA with $P_{\alpha}$ = 0.43. }

\label{fig6}
\end{figure}

\begin{figure}
{
\includegraphics[width=1.5in]{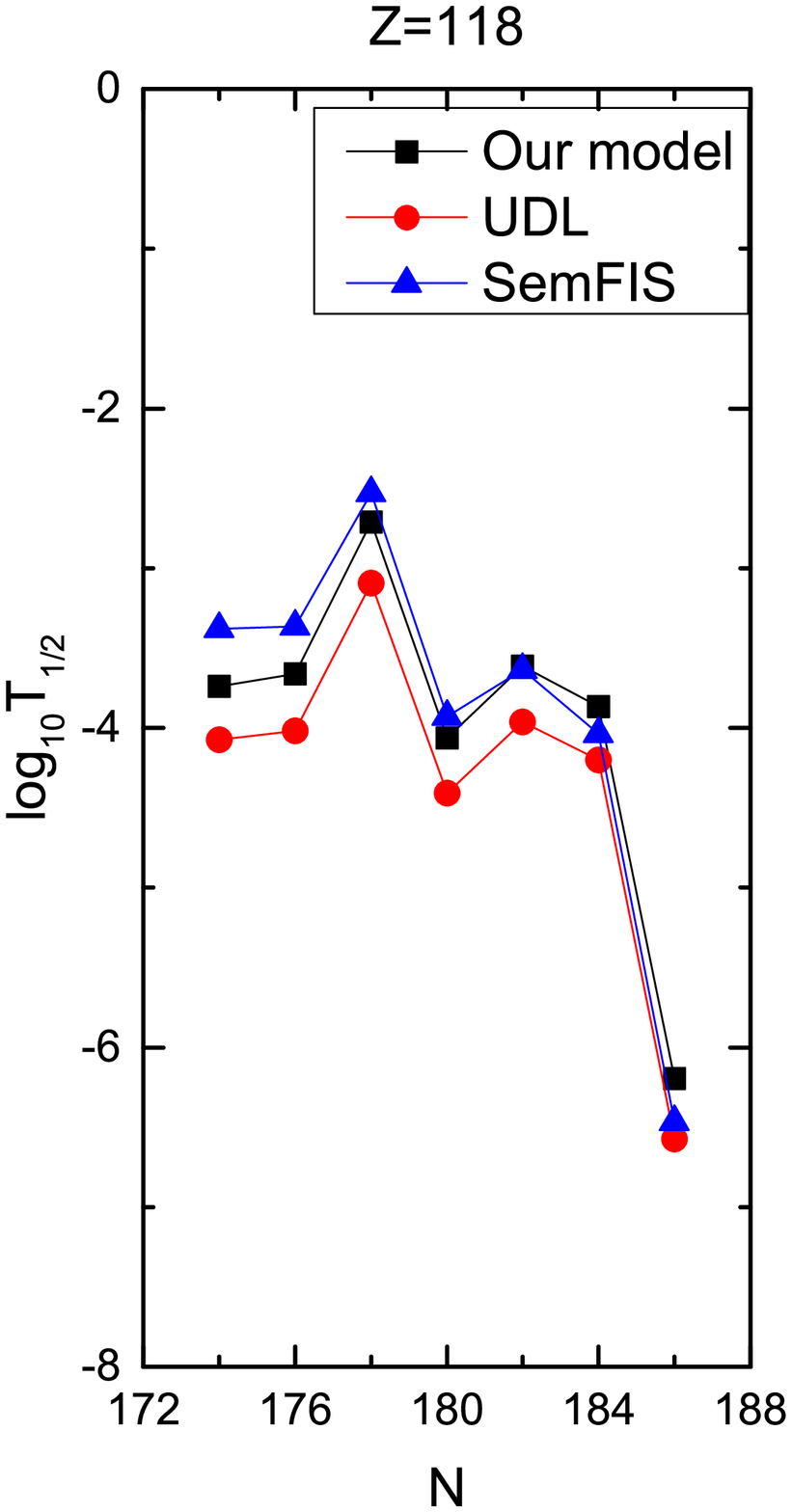}}
\hspace{0in}
{
\includegraphics[width=1.5in]{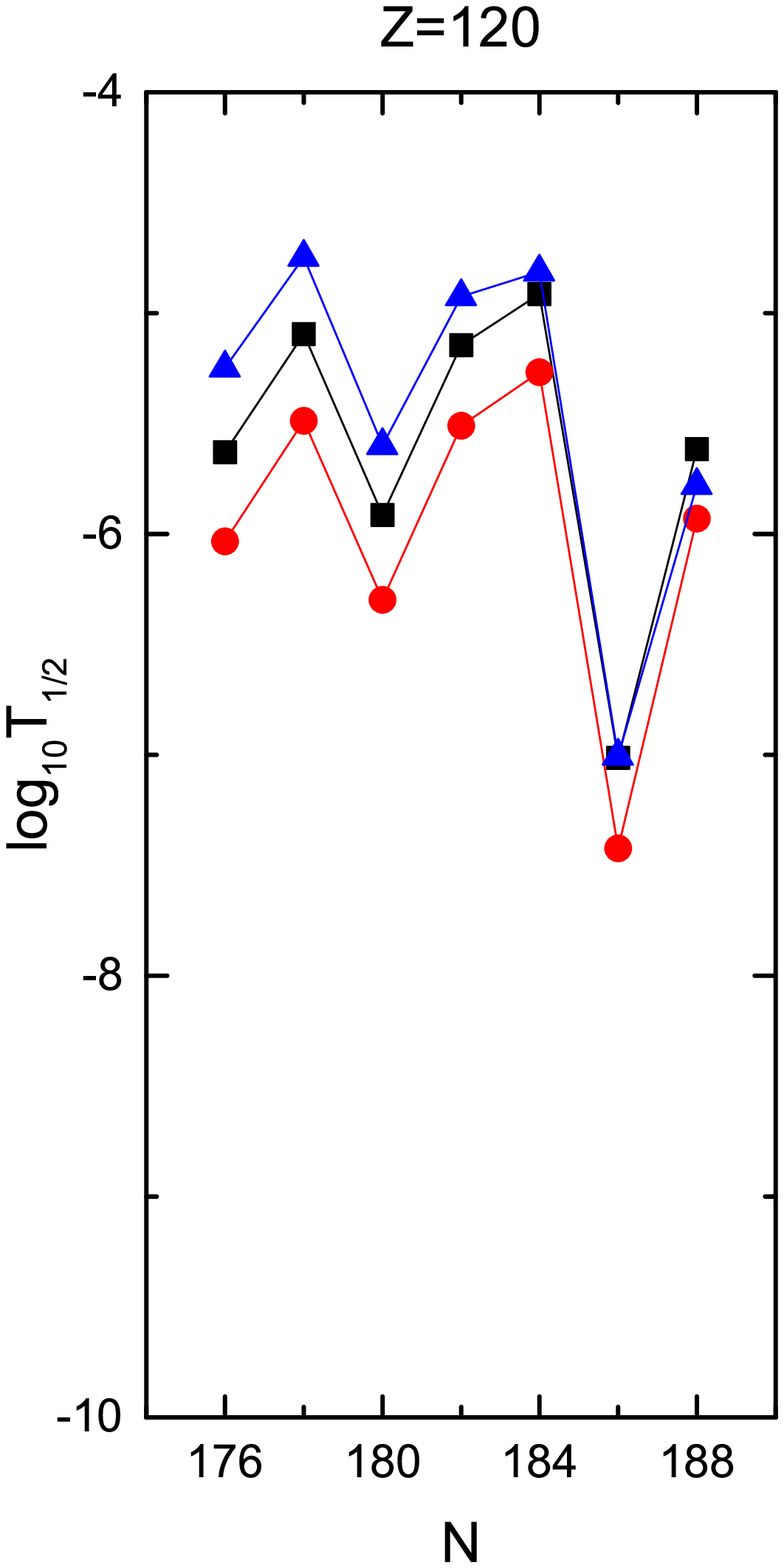}}
\hspace{0in}
{
\includegraphics[width=1.5in]{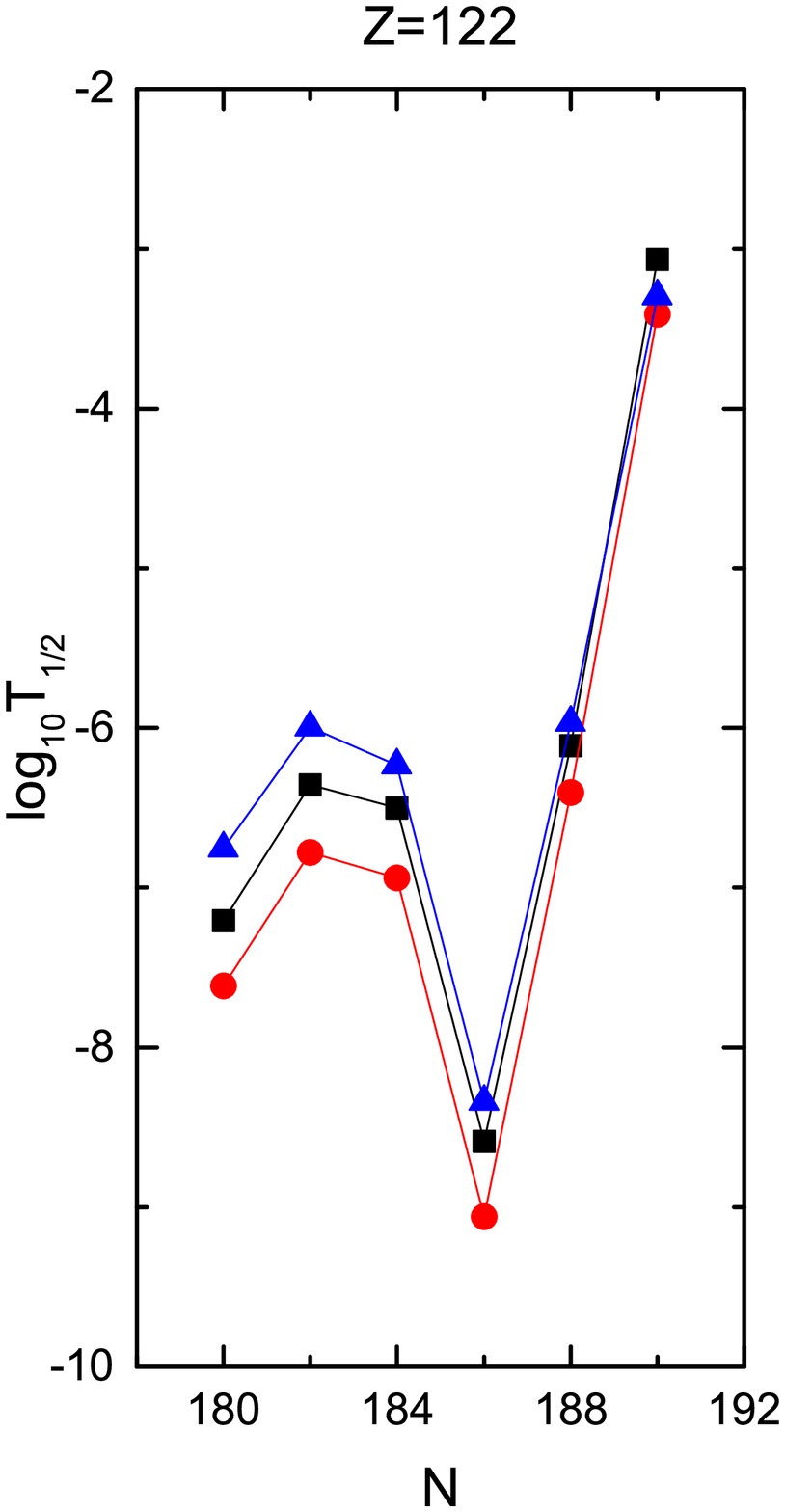}}
\hspace{0in}

{
\includegraphics[width=1.5in]{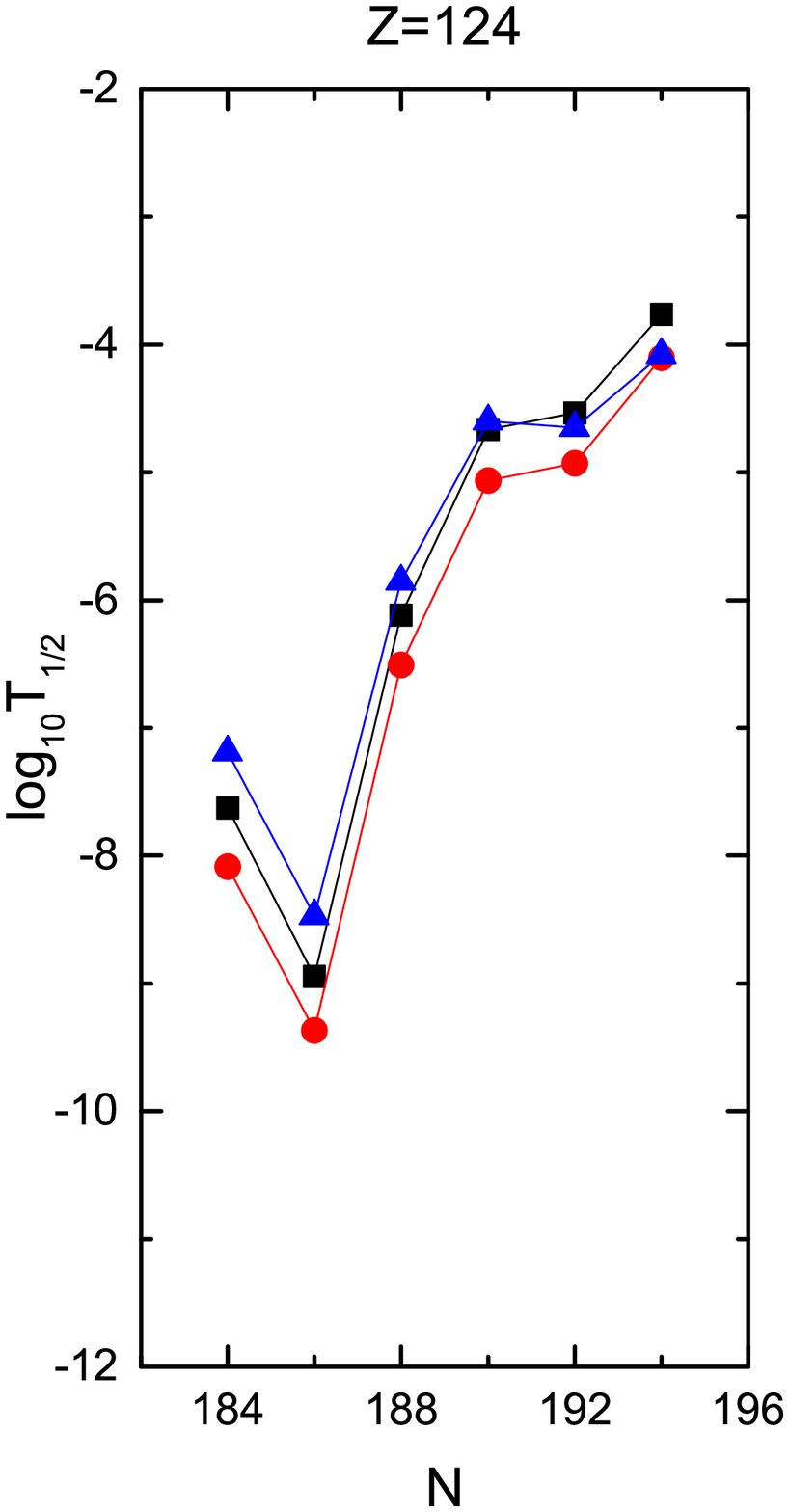}}
\hspace{0in}
{
\includegraphics[width=1.5in]{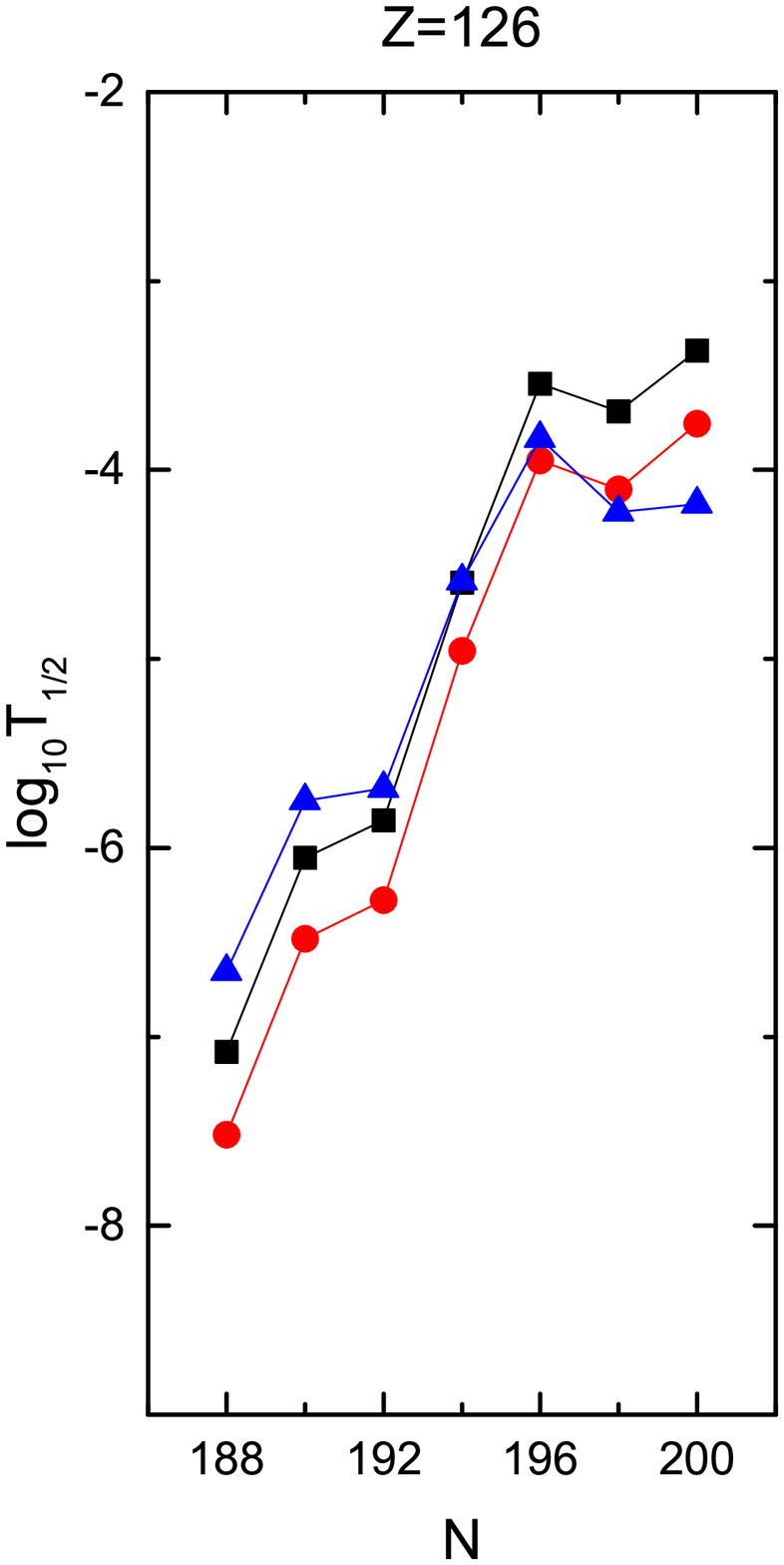}}
\hspace{0in}
{
\includegraphics[width=1.5in]{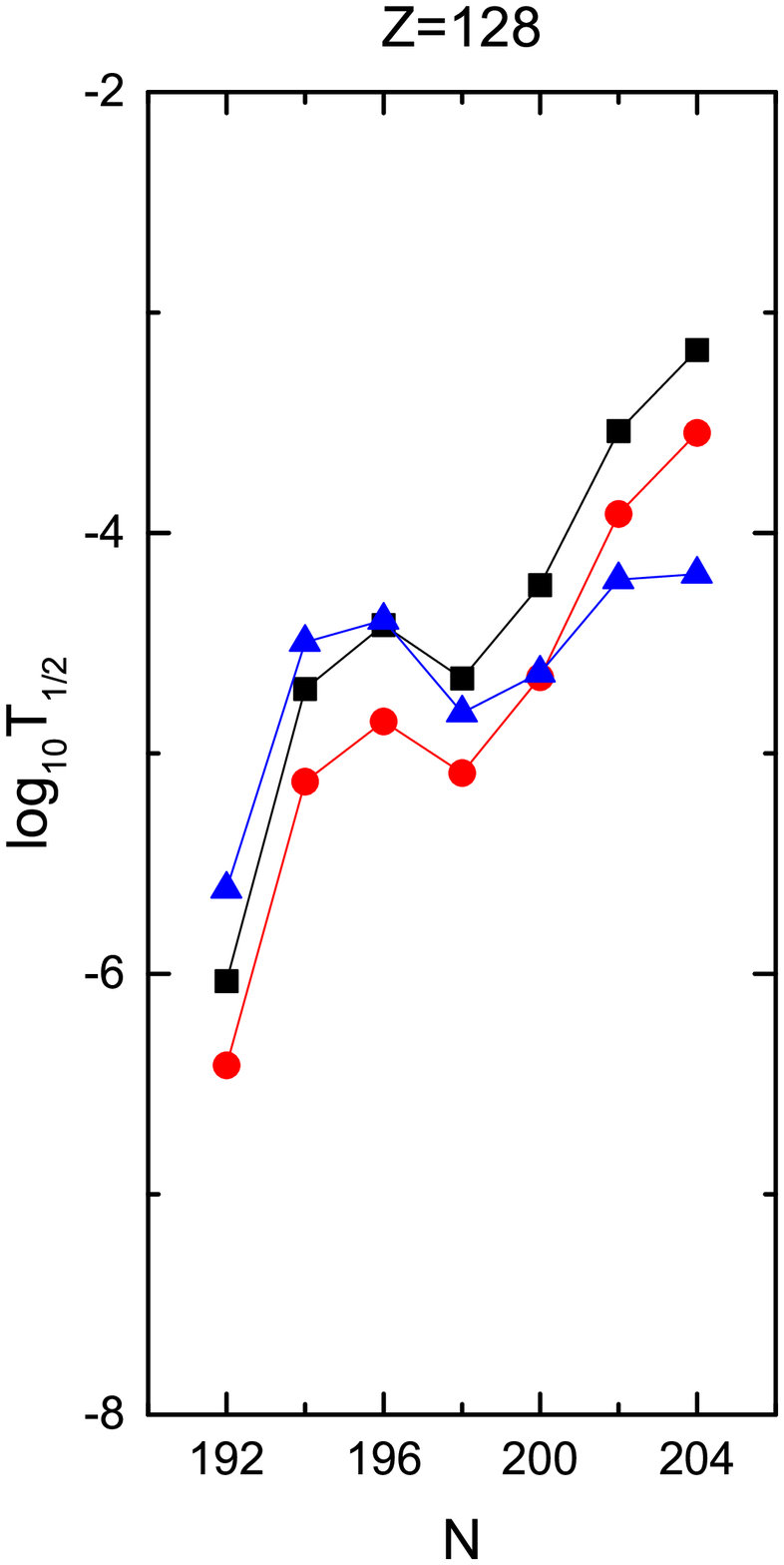}}
\hspace{0in}
\caption{(color online) The predicted $\alpha$ decay half-lives in the logarithmic form using three different models for even--even nuclei with \emph{Z} = 118, 120, 122, 124, 126 and 128 isotopes. The blue triangle, black square and red dot represent the $log_{10}T_{1/2}$ predicted by the SemFIS, our model and UDL, respectively. }
\end{figure}

In the following, we extend our model to predict the $\alpha$ decay half-lives of superheavy nuclei with $104 \leqslant Z \leqslant 128$. For comparing, we also calculate the $\alpha$ decay half-lives of these nuclei by SemFIS and UDL. In the $\alpha$ decay, the half-life $T_{1/2}$ is extremely sensitive to $\alpha$ decay energy $Q_{\alpha}$. Therefore, to obtain the precisely predictions of $\alpha$ decay half-lives for the heavy and superheavy nuclei, the method of selecting a more precise $Q_{\alpha}$ is at the heart of the matter.  Recently, A. Sobiczewski compared nine different mass models: Moller ${et\ al.}$ (FRDM) \cite{73}, Duflo and Zuker (DZ) \cite{74}, Nayak and Satpathy (INM) \cite{75}, Wang and Liu (WS3+) \cite{76}, Wang ${et\ al.}$ (WS4+) \cite{77,78}, Muntian ${et\ al.}$ (HN) \cite{79,80}, Kuzmina ${et\ al.}$ (TCSM) \cite{81}, Goriely ${et\ al.}$ (HFB31) \cite{82} and Liran ${et\ al.}$ (SE) \cite{39}, found that the WS3+ model \cite{76} is the most accurate in reproducing the experimental $Q_{\alpha}$ of superheavy nuclei \cite{34,36}.

Therefore, taking the $\alpha$ decay energy $Q_{\alpha}$ from WS3+ \cite{76}, we systematically predict the $\alpha$ decay half-lives of 64 nuclei with $104 \leqslant Z \leqslant 128$ whose $\alpha$ decay is energetically allowed or observed but not yet quantified with our model, SemFIS and UDL. The detailed results are listed in the Tab. 2. In this table, the first three columns denote mass number of parent nucleus, proton number of parent nucleus and  theoretical $\alpha$ decay energy which is taken from WS3+ \cite{76}, respectively. The fourth column denotes $\alpha$ preformation factor calculated with CFM. The last three columns represent predicted $\alpha$ decay half-life by our model, UDL and SemFIS, which are denoted as  $T_{1/2}^{\rm{pre1}}$, $T_{1/2}^{\rm{pre2}}$ and $T_{1/2}^{\rm{pre3}}$, respectively. From this table, we can clearly see that the prediction results by using three models are basically the same. To intuitively display, we plot the logarithmic form of predicted $\alpha$ decay half-lives for isotopes chains \emph{Z} = 118, 120, 122, 124, 126 and 128 in Fig. 2. In this figure, the blue triangle, black square and red dot represent the predictions by the SemFIS, our model and UDL, respectively. It is found that the three curves have a consistent trend of change and the half-lives calculated by the our model is well in between the ones calculated by the other two models.

\clearpage
\begingroup
\begin{longtable}{ccccccc}
\caption{Predictions of $\alpha$ decay half-lives and the  $\alpha$ preformation factors of even--even nuclei from \emph{Z} = 104 to 128. The $\alpha$ decay energies are taken from WS3+ \cite{76}. The $\alpha$ preformation factors $P_{\alpha}$ are calculated within the CFM\cite{14,31,32,37,38}. }
\label{table 1} \\
\hline
\hline$A$&$Z$&${Q_{\alpha}^{WS3+}\rm{(MeV)}}$&$P_{\alpha}$&$T_{1/2}^{\rm{pre1}\rm{(s)}}$&$T_{1/2}^{\rm{pre2}\rm{(s)}}$&$T_{1/2}^{\rm{pre3}\rm{(s)}}$\\
\hline
\endfirsthead
\multicolumn{6}{c}%
{{\tablename\ \thetable{} -- (\emph{continued}) }} \\
\hline
\hline$A$&$Z$&${Q_{\alpha}^{WS3+}\rm{(MeV)}}$&$P_{\alpha}$&$T_{1/2}^{\rm{pre1}\rm{(s)}}$&$T_{1/2}^{\rm{pre2}\rm{(s)}}$&$T_{1/2}^{\rm{pre3}\rm{(s)}}$\\
\hline
\endhead
\hline
\hline \multicolumn{6}{r}{{(\emph{continued on next page})}} \\
\endfoot
\hline
 \hline
\endlastfoot
$	256	$	&	104	&	 8.960 	&	0.217 	&	$ 1.10\times10^{0}$	    &	$ 6.81\times10^{ 1}$	&	$ 1.15\times10^{0}$	    \\
$	258	$	&	104	& 	 9.241 	&	0.217 	&	$ 1.43\times10^{-1}$	&	$ 9.03\times10^{-2}$	&	$ 1.72\times10^{-1}$	\\
$	260	$	&	104	&	 8.917 	&	0.213 	&	$ 1.26\times10^{0}$	    &	$ 7.92\times10^{-1}$	&	$ 1.55\times10^{0}$ 	\\
$	260	$	&	106	&	 9.940 	&	0.208 	&	$ 8.35\times10^{-3}$	&	$ 4.69\times10^{-3}$	&	$ 9.04\times10^{-3}$	\\
$	264	$	&	108	&	 10.627 &	0.247 	&	$ 5.21\times10^{-4}$	&	$ 3.24\times10^{-4}$	&	$ 7.13\times10^{-4}$	\\
$	268	$	&	108	&	 9.847 	&	0.200 	&	$ 5.59\times10^{-2}$	&	$ 3.00\times10^{-2}$	&	$ 7.71\times10^{-2}$	\\
$	270	$	&	108	&	 9.185 	&	0.203 	&	$ 4.26\times10^{0}$	&	$ 2.38\times10^{ 0}$	&	$ 6.03\times10^{0}$	    \\
$	270	$	&	110	&	10.881 &	0.222 	&	$ 5.10\times10^{-4}$	&	$ 2.73\times10^{-4}$	&	$ 7.67\times10^{-4}$	\\
$	262	$	&	108	&	10.994 &	0.213 	&	$ 8.91\times10^{-5}$	&	$ 4.58\times10^{-5}$	&	$ 8.60\times10^{-5}$	\\
$	272	$	&	108	&	 9.541 	&	0.202 	&	$ 3.39\times10^{-1}$	&	$ 1.90\times10^{-1}$	&	$ 4.23\times10^{-1}$	\\
$	266	$	&	110	&	12.137 &	0.237 	&	$ 1.01\times10^{-6}$	&	$ 4.98\times10^{-7}$	&	$ 1.18\times10^{-6}$	\\
$	276	$	&	110	&	10.977 &	0.222 	&	$ 2.29\times10^{-4}$	&	$ 1.26\times10^{-4}$	&	$ 3.50\times10^{-4}$	\\
$	278	$	&	110	&	10.310 &	0.215 	&	$ 9.72\times10^{-3}$	&	$ 5.51\times10^{-3}$	&	$ 1.23\times10^{-2}$	\\
$	270	$	&	112	&	12.242 &	0.250 	&	$ 2.03\times10^{-6}$	&	$ 1.01\times10^{-6}$	&	$ 2.52\times10^{-6}$	\\
$	280	$	&	112	&	10.912 &	0.252 	&	$ 1.13\times10^{-3}$	&	$ 6.88\times10^{-4}$	&	$ 2.17\times10^{-3}$	\\
$	282	$	&	112	&	10.106 &	0.206 	&	$ 1.53\times10^{-1}$	&	$ 8.05\times10^{-2}$	&	$ 2.11\times10^{-1}$	\\
$	284	$	&	114	&	10.666 &	0.257 	&	$ 1.82\times10^{-2}$	&	$ 1.11\times10^{-2}$	&	$ 4.03\times10^{-2}$	\\
$	286	$	&	114	&	 9.944 	&	0.219 	&	$ 1.79\times10^{0}$	    &	$ 9.69\times10^{-1}$	&	$ 3.07\times10^{0}$	    \\
$	288	$	&	114	&	 9.473 	&	0.197 	&	$ 4.68\times10^{1}$	    &	$ 2.31\times10^{1}$	    &	$ 5.77\times10^{1}$	    \\
$	288	$	&	116	&	11.105 	&	0.206 	&	$ 7.10\times10^{-3}$	&	$ 3.28\times10^{-3}$	&	$ 1.39\times10^{-2}$	\\
$	290	$	&	116	&	10.878 &	0.204 	&	$ 2.43\times10^{-2}$	&	$ 1.14\times10^{-2}$	&	$ 4.24\times10^{-2}$	\\
$	292	$	&	116	&	10.917 &	0.184 	&	$ 1.95\times10^{-2}$	&	$ 8.34\times10^{-3}$	&	$ 2.52\times10^{-2}$	\\
$	294	$	&	116	&	10.451 &	0.185 	&	$ 2.94\times10^{-1}$	&	$ 1.31\times10^{-1}$	&	$ 2.86\times10^{-1}$	\\
$	296	$	&	116	&	10.778 &	0.184 	&	$ 3.72\times10^{-2}$	&	$ 1.64\times10^{-2}$	&	$ 2.49\times10^{-2}$	\\
$	292	$	&	118	&	12.015 &	0.230 	&	$ 1.81\times10^{-4}$	&	$ 8.40\times10^{-5}$	&	$ 4.18\times10^{-4}$	\\
$	294	$	&	118	&	11.974 	&	0.216 	&	$ 2.18\times10^{-4}$	&	$ 9.59\times10^{-5}$	&	$ 4.32\times10^{-4}$	\\
$	296	$	&	118	&	11.562 	&	0.193 	&	$ 1.95\times10^{-3}$	&	$ 8.05\times10^{-4}$	&	$ 2.96\times10^{-3}$	\\
$	298	$	&	118	&	12.118 	&	0.222 	&	$ 8.62\times10^{-5}$	&	$ 3.92\times10^{-5}$	&	$ 1.18\times10^{-4}$	\\
$	300	$	&	118	&	11.905 	&	0.213 	&	$ 2.44\times10^{-4}$	&	$ 1.09\times10^{-4}$	&	$ 2.31\times10^{-4}$	\\
$	302	$	&	118	&	11.995 	&	0.218 	&	$ 1.37\times10^{-4}$	&	$ 6.32\times10^{-5}$	&	$ 9.18\times10^{-5}$	\\
$	304	$	&	118	&	13.104 &	0.228 	&	$ 6.36\times10^{-7}$	&	$ 2.68\times10^{-7}$	&	$ 3.42\times10^{-7}$	\\
$	296	$	&	120	&	13.188 &	0.233 	&	$ 2.34\times10^{-6}$	&	$ 9.25\times10^{-7}$	&	$ 5.65\times10^{-6}$	\\
$	298	$	&	120	&	12.900 &	0.227 	&	$ 8.06\times10^{-6}$	&	$ 3.26\times10^{-6}$	&	$ 1.80\times10^{-5}$	\\
$	300	$	&	120	&	13.287 &	0.242 	&	$ 1.22\times10^{-6}$	&	$ 5.02\times10^{-7}$	&	$ 2.52\times10^{-6}$	\\
$	302	$	&	120	&	12.878 &	0.237 	&	$ 7.17\times10^{-6}$	&	$ 3.10\times10^{-6}$	&	$ 1.19\times10^{-5}$	\\
$	304	$	&	120	&	12.745 &	0.237 	&	$ 1.22\times10^{-5}$	&	$ 5.42\times10^{-6}$	&	$ 1.54\times10^{-5}$	\\
$	306	$	&	120	&	13.823 &	0.245 	&	$ 9.66\times10^{-8}$	&	$ 3.78\times10^{-8}$	&	$ 9.85\times10^{-8}$	\\
$	308	$	&	120	&	13.036 &	0.265 	&	$ 2.42\times10^{-6}$	&	$ 1.17\times10^{-6}$	&	$ 1.66\times10^{-6}$	\\
$	302	$	&	122	&	14.262 &	0.274 	&	$ 6.21\times10^{-8}$	&	$ 2.43\times10^{-8}$	&	$ 1.76\times10^{-7}$	\\
$	304	$	&	122	&	13.791 &	0.244 	&	$ 4.38\times10^{-7}$	&	$ 1.67\times10^{-7}$	&	$ 1.00\times10^{-6}$	\\
$	306	$	&	122	&	13.859 &	0.236 	&	$ 3.15\times10^{-7}$	&	$ 1.15\times10^{-7}$	&	$ 5.83\times10^{-7}$	\\
$	308	$	&	122	&	15.040 &	0.268 	&	$ 2.57\times10^{-9}$	&	$ 8.68\times10^{-10}$	&	$ 4.56\times10^{-9}$	\\
$	310	$	&	122	&	13.543 &	0.309 	&	$ 7.71\times10^{-7}$	&	$ 3.94\times10^{-7}$	&	$ 1.08\times10^{-6}$	\\
$	312	$	&	122	&	12.112 	&  0.226 	&	$ 8.60\times10^{-4}$	&	$ 3.87\times10^{-4}$	&	$ 5.08\times10^{-4}$	\\
$	308	$	&	124	&	14.771 &	0.268 	&	$ 2.37\times10^{-8}$	&	$ 8.23\times10^{-9}$	&	$ 6.42\times10^{-8}$	\\
$	310	$	&	124	&	15.517 &	0.338 	&	$ 1.13\times10^{-9}$	&	$ 4.30\times10^{-10}$	&	$ 3.36\times10^{-9}$	\\
$	312	$	&	124	&	13.868 &	0.266 	&	$ 7.58\times10^{-7}$	&	$ 3.10\times10^{-7}$	&	$ 1.40\times10^{-6}$	\\
$	314	$	&	124	&	13.124 &	0.234 	&	$ 2.16\times10^{-5}$	&	$ 8.67\times10^{-6}$	&	$ 2.51\times10^{-5}$	\\
$	316	$	&	124	&	13.044 &	0.229 	&	$ 2.93\times10^{-5}$	&	$ 1.18\times10^{-5}$	&	$ 2.25\times10^{-5}$	\\
$	318	$	&	124	&	12.644 &	0.247 	&	$ 1.71\times10^{-4}$	&	$ 7.86\times10^{-5}$	&	$ 8.32\times10^{-5}$	\\
$	314	$	&	126	&	14.690 &	0.279 	&	$ 8.31\times10^{-8}$	&	$ 3.04\times10^{-8}$	&	$ 2.23\times10^{-7}$	\\
$	316	$	&	126	&	14.107 &	0.256 	&	$ 8.88\times10^{-7}$	&	$ 3.30\times10^{-7}$	&	$ 1.77\times10^{-6}$	\\
$	318	$	&	126	&	13.983 &	0.253 	&	$ 1.40\times10^{-6}$	&	$ 5.28\times10^{-7}$	&	$ 2.07\times10^{-6}$	\\
$	320	$	&	126	&	13.303 &	0.260 	&	$ 2.54\times10^{-5}$	&	$ 1.10\times10^{-5}$	&	$ 2.59\times10^{-5}$	\\
$	322	$	&	126	&	12.812 &	0.219 	&	$ 2.87\times10^{-4}$	&	$ 1.12\times10^{-4}$	&	$ 1.47\times10^{-4}$	\\
$	324	$	&	126	&	12.868 &	0.215 	&	$ 2.04\times10^{-4}$	&	$ 7.89\times10^{-5}$	&	$ 5.98\times10^{-5}$	\\
$	326	$	&	126	&	12.694 &	0.222 	&	$ 4.29\times10^{-4}$	&	$ 1.76\times10^{-4}$	&	$ 6.62\times10^{-5}$	\\
$	320	$	&	128	&	14.326 &	0.299 	&	$ 9.27\times10^{-7}$	&	$ 3.84\times10^{-7}$	&	$ 2.43\times10^{-6}$	\\
$	322	$	&	128	&	13.648 &	0.245 	&	$ 1.96\times10^{-5}$	&	$ 7.43\times10^{-6}$	&	$ 3.19\times10^{-5}$	\\
$	324	$	&	128	&	13.497	&	0.233 	&	$ 3.81\times10^{-5}$	&	$ 1.40\times10^{-5}$	&	$ 4.01\times10^{-5}$	\\
$	326	$	&	128	&	13.596 &	0.235 	&	$ 2.19\times10^{-5}$	&	$ 8.14\times10^{-6}$	&	$ 1.53\times10^{-5}$	\\
$	328	$	&	128	&	13.369 &	0.232 	&	$ 5.78\times10^{-5}$	&	$ 2.21\times10^{-5}$	&	$ 2.31\times10^{-5}$	\\
$	330	$	&	128	&	13.005 &	0.242 	&	$ 2.91\times10^{-4}$	&	$ 1.22\times10^{-4}$	&	$ 6.14\times10^{-5}$	\\
$	332	$	&	128	&	12.823 &	0.236 	& $ 6.78 \times10^{-4}$   &	$ 2.84\times10^{-4}$	&	$ 6.51\times10^{-5}$	 \\
\end{longtable}
\endgroup
Predicting new magic numbers has been an interesting topic in nuclear physics. To observe possible magic numbers, we plot the changes of $\alpha$ decay energies $Q_{\alpha}$ values and the $\alpha$ preformation factors $P_\alpha$ of \emph{Z} = 118, 120, 122, 124, 126 and 128 isotopes with increasing neutron number \emph{N} in  Fig. 3. The $Q_{\alpha}$ values are taken from WS3+ \cite{76}. From this figure, we can clearly see that:

(1) For the \emph{Z} = 118 isotope, the $Q_{\alpha}$ and $P_\alpha$ decrease until \emph{N} = 178, increase from \emph{N} = 178 to 180, and decrease again until \emph{N} = 182. Then, the $Q_{\alpha}$ and $P_\alpha$ increase with the increasing of neutron number up to \emph{N} = 186.

(2) For the \emph{Z} = 120 isotope, the $Q_{\alpha}$ and $P_\alpha$ also decrease with the increasing of neutron number until \emph{N} = 178, then begin to increase from \emph{N} = 178 to 180, and then reduce again with \emph{N} until \emph{N} = 184. In the end, the $Q_{\alpha}$ and $P_\alpha$ decrease swiftly with \emph{N}.

 (3) For the \emph{Z} = 122 isotope, the $Q_{\alpha}$ drop with the increasing of neutron number \emph{N} until \emph{N} = 182. Then, the $Q_{\alpha}$ rise with the increasing of neutron number \emph{N} and reach a maximum when \emph{N} = 186. Finally, the $Q_{\alpha}$ drop rapidly with the neutron number \emph{N}. The trend of $P_\alpha$ with \emph{N} is similar to that of $Q_{\alpha}$ with \emph{N}, the difference is that  $P_\alpha$ gets a minimum value when \emph{N} = 184, and a maximum when \emph{N} = 188.

(4) For the \emph{Z} = 124 isotope, the $Q_{\alpha}$ and $P_\alpha$ all become larger as \emph{N} increases until \emph{N} = 186, then the $Q_{\alpha}$ begin to sharply decrease, the $P_\alpha$ increase with \emph{N} and then become  larger.

 (5) For the \emph{Z} = 126 and 128 isotopes, the changing trend of  $Q_{\alpha}$ with the increasing of neutron number \emph{N} is basically the same. At first, the $Q_{\alpha}$ fall sharply with the increasing of neutron number \emph{N} until the neutron number  \emph{N} = 196. Then, the $Q_{\alpha}$ rise with the increasing of neutron number \emph{N} until the neutron number \emph{N} = 198. In the end, the $Q_{\alpha}$ fall again with the  increasing of neutron number. The changing trend of $P_\alpha$ of \emph{Z} = 126 isotope with the increasing of neutron number is similar to that  of \emph{Z} = 118 isotope $P_\alpha$ with the increasing of neutron number \emph{N}. The difference is that for \emph{Z} = 126 isotope, the maximum value of $P_\alpha$ is at  \emph{N} = 194, while the minimum values are at  \emph{N} = 192 and 198. For \emph{Z} = 128 isotope, the trend of $P_\alpha$  with the increasing of neutron number \emph{N} is: sharply decrease, then slowly increase, slowly decrease, increase again and then decrease again.

Combined with Fig. 3. and the above brief analysis, we find that for the \emph{Z} = 118 and 120 isotopes, whether it is the trend of $Q_{\alpha}$ with \emph{N} or the trend of $P_\alpha$ with \emph{N}, \emph{N} = 178 is always the minimum point of $Q_{\alpha}$ and $P_\alpha$. It implies that neutron number \emph{N} = 178 may be the neutron magic number. For the \emph{Z} = 122, 124, 126 and 128 isotopes, we compared the changing trend of $Q_{\alpha}$ and $P_\alpha$ with \emph{N}, we don't find a common maximum or minimum value.

\begin{figure}
%\centering
\begin{minipage}[c]{0.5\textwidth}
%\centering
\includegraphics[height=6cm,width=8cm]{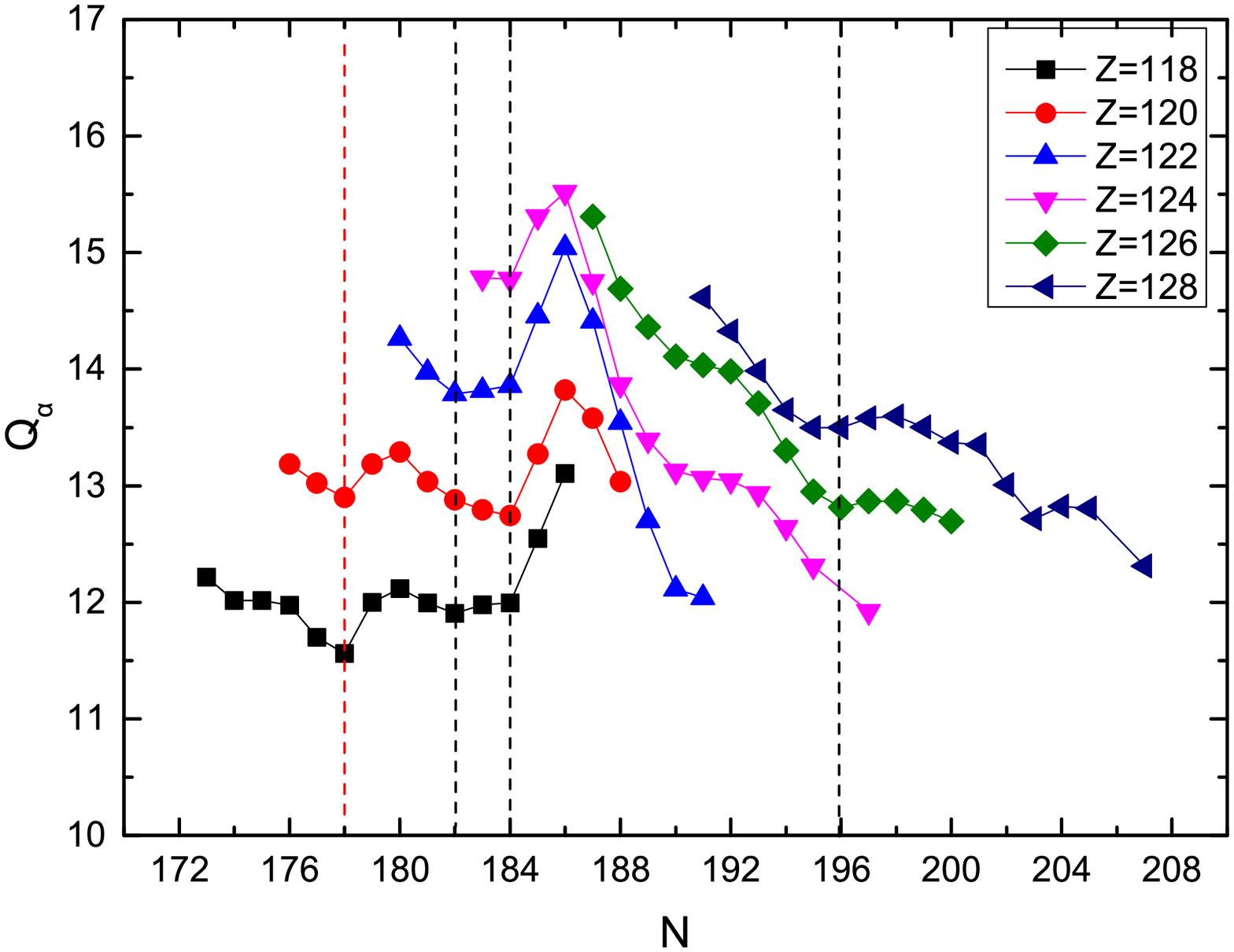}
\end{minipage}%
\begin{minipage}[c]{0.5\textwidth}
%\centering
\includegraphics[height=6cm,width=8cm]{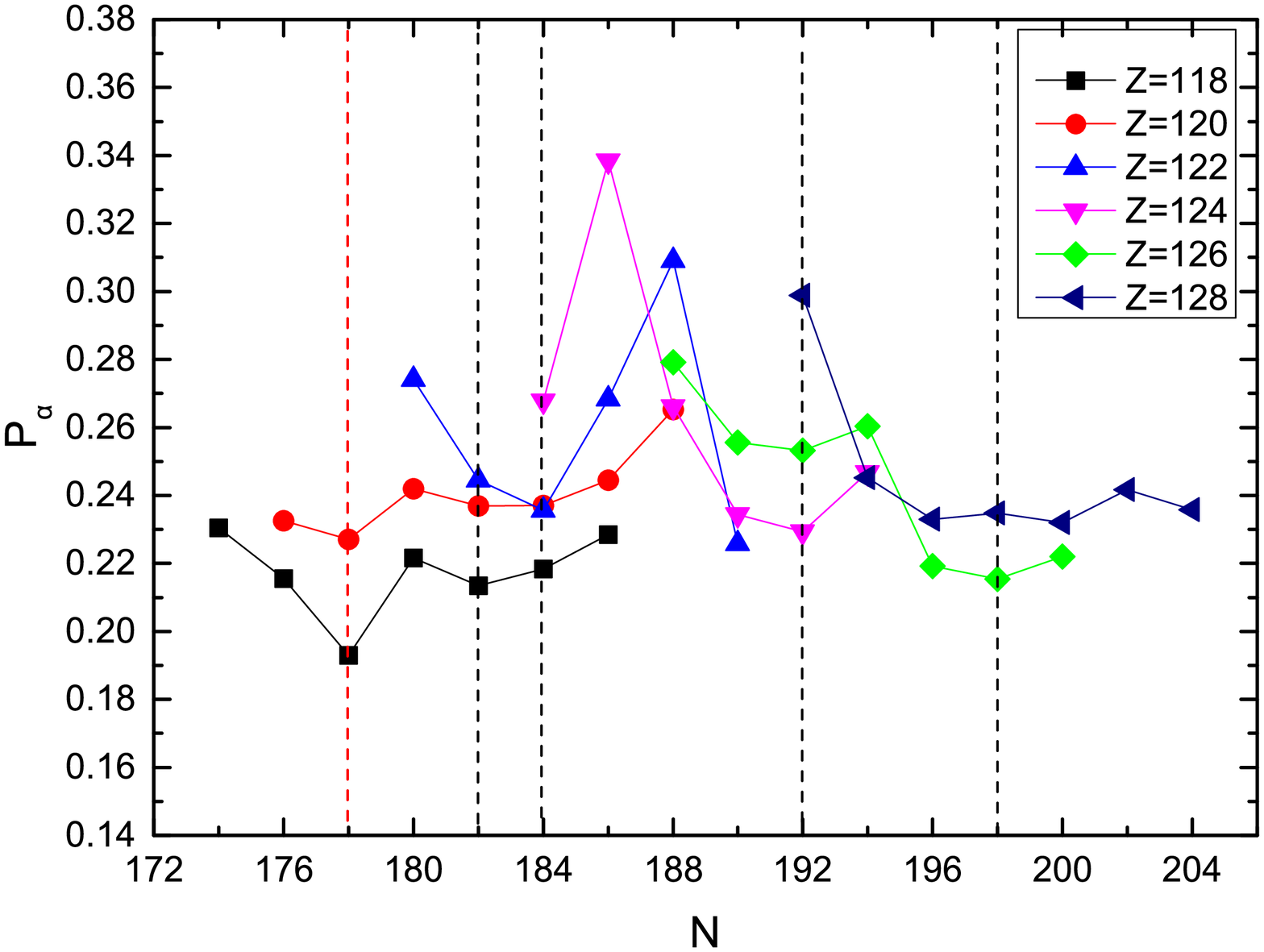}
\end{minipage}
\caption{ (color online) The $\alpha$ decay energies $Q_{\alpha}$ (left-hand side) and the $\alpha$ preformation factors $P_\alpha$ (right-hand side) with neutron number \emph{N} for \emph{Z} = 118, 120, 122, 124, 126 and 128 isotopes.}
\end{figure}
\section{Summary}
  In this work, we use TPA taking $P_\alpha$ within CFM systematically calculate the $\alpha$ decay half-lives of 170 even--even nuclei with $60 \leqslant Z \leqslant 118$. The calculated results can well reproduce the experimental data. For benchmark, we also calculate the $\alpha$ decay half-lives of these nuclei using TPA taking $P_{\alpha}$ = 0.43, SemFIS and UDL. The calculations of  these models are basically consistent. Then, we predict the $\alpha$ decay half-lives of the SHN with $104 \leqslant Z \leqslant 128$ within our model, SemFIS and UDL, while the $\alpha$ decay energies are taken from WS3+. The predictions of the three models are basically the same. Meanwhile, we systematically study the changes of $Q_{\alpha}$  and $P_\alpha$ values of \emph{Z} = 118, 120, 122, 124, 126 and 128 isotopes with increasing neutron number \emph{N}. Based on the results, we infer that \emph{N} = 178 may be the neutron magic number. This work will be used as a reference for experimental and theoretical research in the future.

\section*{Acknowledgements}

This work is supported in part by the National Natural Science Foundation of China (Grants No. 11205083 and No.11505100), the Construct Program of the Key Discipline in Hunan Province, the Research Foundation of Education Bureau of Hunan Province, China (Grant No. 15A159 and No.18A237), the Natural Science Foundation of Hunan Province, China (Grants No. 2015JJ3103 and No. 2015JJ2121), the Innovation Group of Nuclear and Particle Physics in USC, the Shandong Province Natural Science Foundation, China (Grant No. ZR2015AQ007), the Hunan Provincial Innovation Foundation For Postgraduate (Grant No. CX20190714), the National Innovation Training Foundation of China (Grant No.201910555161), and the Opening Project of Cooperative Innovation Center for Nuclear Fuel Cycle Technology and Equipment, University of South China (Grant No. 2019KFZ10).

\appendix

\end{document}